\documentclass[sigconf]{acmart}
\def\BibTeX{{\rm B\kern-.05em{\sc i\kern-.025em b}\kern-.08emT\kern-.1667em\lower.7ex\hbox{E}\kern-.125emX}}
    
\usepackage{comment}
\usepackage{url}
\usepackage{booktabs}
\usepackage[caption=false,font=footnotesize,labelfont=sf,textfont=sf]{subfig}
\usepackage{tcolorbox}

\newcommand{\RqOne}{\textbf{(RQ1):} \emph{What are the characteristics of software supply chain maps?}}
\newcommand{\RqTwo}{\textbf{(RQ2):} \emph{How have software supply chain maps evolved?}}
\newcommand{\RqThree}{\textbf{(RQ3):} \emph{How source files are transferred in software supply chain maps?}}

\begin{document}

%
\title{Software Supply Chain Map: How Reuse Networks Expand}

%
\author{Hideaki Hata}
\affiliation{%
  \institution{Shinshu University}
}
\email{hata@shinshu-u.ac.jp}

\author{Takashi Ishio}
\affiliation{%
  \institution{Nara Institute of Science and Technology}
}
\email{ishio@is.naist.jp}

%

%
\begin{abstract}
Clone-and-own is a typical code reuse approach because of its simplicity and efficiency. Cloned software components are maintained independently by a new owner. These clone-and-own operations can be occurred sequentially, that is, cloned components can be cloned again and owned by other new owners on the supply chain. In general, code reuse is not documented well, consequently, appropriate changes like security patches cannot be propagated to descendant software projects. However, the OpenChain Project defined identifying and tracking source code reuses as responsibilities of FLOSS software staffs. Hence supporting source code reuse awareness is in a real need. This paper studies software reuse relations in FLOSS ecosystem. Technically, clone-and-own reuses of source code can be identified by file-level clone set detection. Since change histories are associated with files, we can determine origins and destinations in reusing across multiple software by considering times. By building software supply chain maps, we find that clone-and-own is prevalent in FLOSS development, and set of files are reused widely and repeatedly. These observations open up future challenges of maintaining and tracking global software genealogies.
\end{abstract}

%
%

\ccsdesc[500]{Software and its engineering~Software libraries and repositories}
\ccsdesc[500]{Software and its engineering~Reusability}
\ccsdesc[500]{Software and its engineering~Software evolution}
\ccsdesc[300]{Software and its engineering~Open source model}

\keywords{software supply chain map, clone-and-own, code reuse, FLOSS ecosystem, global version control}

%

%
\maketitle

\section{Introduction}

Because of the nature of free and open source, and the lack of the view of the whole picture, there are several problems in free/libre and open source software (FLOSS), such as
\textit{broken library dependencies}
~\cite{Kula2017} and 
\textit{license violations}
~\cite{7476659}. 
One reason of such issues is in ad hoc code reusing.
Copying software components and then maintaining them by a new owner is one type of code reuse, which is known as \textit{clone-and-own}
~\cite{Dubinsky:2013:ESC:2495256.2495759,Ishio:2017:SFS:3104188.3104222}.
In software product line engineering, the usage of this clone-and-own approach is discouraged, since it makes maintaining multiple product lines difficult. If changes are made in the original or copied components, they are not easily propagated. This is because of the ad hoc nature of component reusing. Developers working in different product lines do not know when clone-and-own operations occurred and where the cloned components are located. Although there are disadvantages, the clone-and-own approach is adopted in FLOSS projects as well as industrial software product lines due to its benefits, such as simplicity, availability, and independence of developers~\cite{Dubinsky:2013:ESC:2495256.2495759}.
Similar to industrial cases, changes are not appropriately propagated 
because of unrecorded and ill-managed clone-and-own reuses.

Even with its large-scale code resources and a large amount of developers, Google is reported to address reusing properly 
with a monolithic source code management system~\cite{Potvin:2016:WGS:2963119.2854146}.
Such a monolithic system has several advantages: unified versioning, extensive code sharing and reuse, simplified dependency management, 
large-scale refactoring, 
and so on~\cite{Potvin:2016:WGS:2963119.2854146}.
Based on the monolithic source code management, changes to core libraries are promptly and easily propagated through the dependency chain into the final products that rely on the libraries~\cite{Potvin:2016:WGS:2963119.2854146}.

Although introducing such a monolithic source code management system in the entire FLOSS ecosystem is not practical, tracking source code across multiple projects is recently considered to be an obligation.
The OpenChain, a Linux Foundation initiative, is working for identifying FLOSS process quality criteria\footnote{\url{https://www.openchainproject.org/}}.
A core set of requirements every quality compliance program must satisfy is defined in the OpenChain Specification~\cite{OpenChainSpec12}. 
As described in the specification, identifying, recording and/or tracking of FLOSS components through reusing source code are considered to be parts of FLOSS responsibilities. Identifying and tracking such source code reuses are required from such an emerging real need.


Instead of preparing a monolithic system, we try constructing networks of clone-and-own relations. We call them \textit{software supply chain maps}.
Source code reuses across projects have been analyzed by code clone detection tools. 
Ossher et~al.~\cite{6080795}, Koschke et~al.~\cite{KoschkeIWSC2016}, and Lopes et~al.~\cite{Lopes:2017:DMC:3152284.3133908} reported that a large number of projects included copies of libraries.
Gharehyazie et~al.~\cite{Gharehyazie:2017:HCC:3104188.3104225} reported that most projects obtain files from outside than provide them.
We extend such analyses to consider the directions of source file reuses across repositories.
To recover file reuse relations across projects, we employ a file-level origin analysis proposed by Ishio et~al.~\cite{Ishio:2017:SFS:3104188.3104222}.
The method identifies the most similar version of sets of files among projects as origins, 
while existing code clone detection techniques simply report all the cloned files in the projects.
By connecting most likely clone-and-own sets of files across projects with considering time information of files, we build software supply chain maps, which represent software reuse relations across software projects.

This paper presents our approach of building software supply chain maps, and conducts an empirical study of clone-and-own activities in FLOSS projects. We consider that this study is important to (1) understand how software reuses spread, (2) understand reuse relations among different software projects, and (3) help developers to maintain their software by applying appropriate changes and propagating changes along with related software projects.








The contributions of this paper are summarized as follows.
\begin{itemize}
\item Proposing the concept of software supply chain maps and designing algorithm to build them.
\item Conducting an empirical study of software supply chain maps with 4,592 C, C++, and Java projects on GitHub. To the best of our knowledge, this is the first study to show directed networks of FLOSS software projects in terms of file reusing.
\item Providing a new insight of source code histories and reuses, such as ``single repositories are not self-contained''. Because files are manually clone-and-owned in newer repositories, the information of original authors and previous histories are lost. This fact can have impact on research related to code repository mining, developer activity analyses (social networks, turnover, etc.), code authorship and ownership.
\end{itemize}

\section{Model of Software Supply Chain Maps}

\begin{figure}
\centering
\includegraphics[width=.9\linewidth]{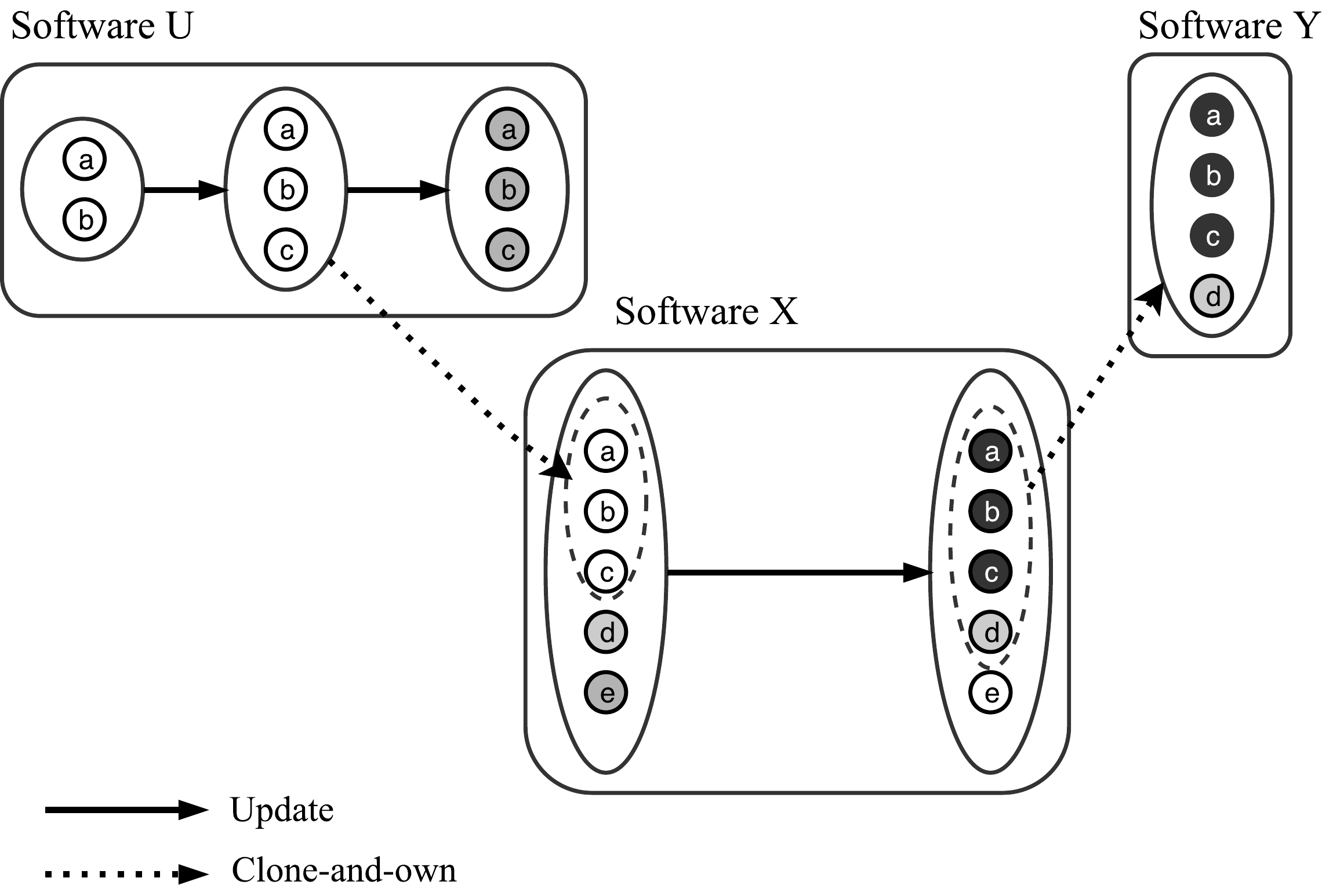}
\caption{An example of a software supply chain lineage.}
\label{fig:lineage}
\end{figure}

\begin{figure}
\includegraphics[width=.9\linewidth]{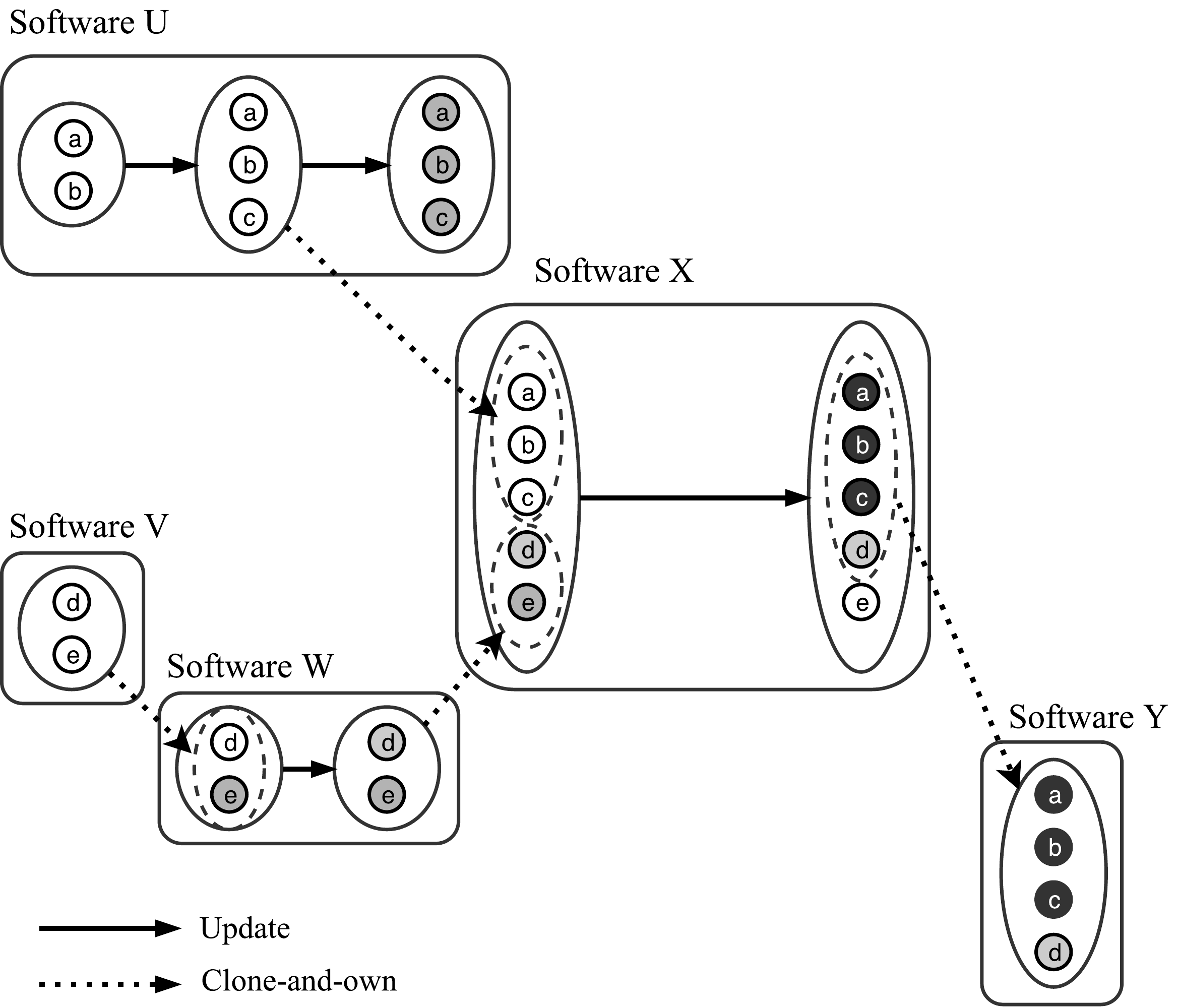}
\caption{An example of a software supply chain map.}
\label{fig:genealogy}
\end{figure}

Similar to clone genealogy~\cite{Kim:2005:ESC:1081706.1081737}, we define a model of software supply chain maps. The genealogies of software reuses describe how software components, sets of source files, are reused and transferred over multiple software projects by means of clone-and-own reusing, and can be regarded as \textit{imaginary source file version transitions across multiple software projects}.

A \textbf{software supply chain lineage} is a directed acyclic graph of snapshots that are related to a specific set of files. It represents the evolution history of a group of files across multiple software projects.
Figure~\ref{fig:lineage} illustrates an example of a clone lineage across software U, X, and Y. In each software, denoted by rounded rectangles, snapshots (files at a specific time or a version) are shown as ellipse. Snapshots are linked with solid arrows from older to newer, which represent updates within software. In software X, there are two snapshots and both include files a, b, c, d, and e (small circles). Different colors for the same files represent the modifications of those files. We see that files of a, b, c, and e are updated from the first version to the second version.
Files a, b, and c of the first version of software X came from software U, more specifically, the second version of software U was reused in software X.
Software Y clone-and-own the file set of a, b, c, and d in the second version of software X. 
Clone-and-own relations are shown as dotted arrows from \textit{origins} (original set of source files) to \textit{destinations} (reused set of source files).

A \textbf{software supply chain map} is a set of software supply chain lineages that are connected with clone-and-own links.
It illustrates how software ingredients~\cite{IshioMSR2016} mixed and transferred to other software projects. 
Figure~\ref{fig:genealogy} shows an example of a software supply chain map, which is extended from Figure~\ref{fig:lineage} by including another clone lineage.
In addition to a clone lineage of software U, X, and Y related to files a, b, c, and d,
the set of files d and e is cloned from software V to software W, and the updated set of files in software W is reused in software X. As a result, all software U, V, W, X, and Y are connected with clone-and-own links in one chain map of a directed network.

\section{Building Software Supply Chain Maps}

Although clone-and-own operations should have been conducted by developers intentionally, they are not always recorded precisely and properly~\cite{XiaJSSST2013}. Therefore, we need to identify those operations and hidden links from multiple software version histories.
As previous studies reported, there are a considerable amount of duplication in source code among multiple projects~\cite{Lopes:2017:DMC:3152284.3133908,Gharehyazie:2017:HCC:3104188.3104225}.
Even in Figure~\ref{fig:lineage} of a clone-and-own lineage, files a, b, and c in the last snapshots of software U, X, and Y can be considered to be a clone group. Therefore, simply detecting similar files is not appropriate to construct software supply chain maps. We need to consider time differences to connect those similar sets with directed arrows. 
Since the previous approach work only on \textit{one-to-many} relations (given a snapshot, identify origins)~\cite{Ishio:2017:SFS:3104188.3104222}, we need to extend their approach to extract whole genealogies with \textit{many-to-many} relations.



\begin{figure*}[!t]
\centering
\subfloat[C2: largest file sets]{
\begin{tabular}{c}
\includegraphics[width=0.17\linewidth]{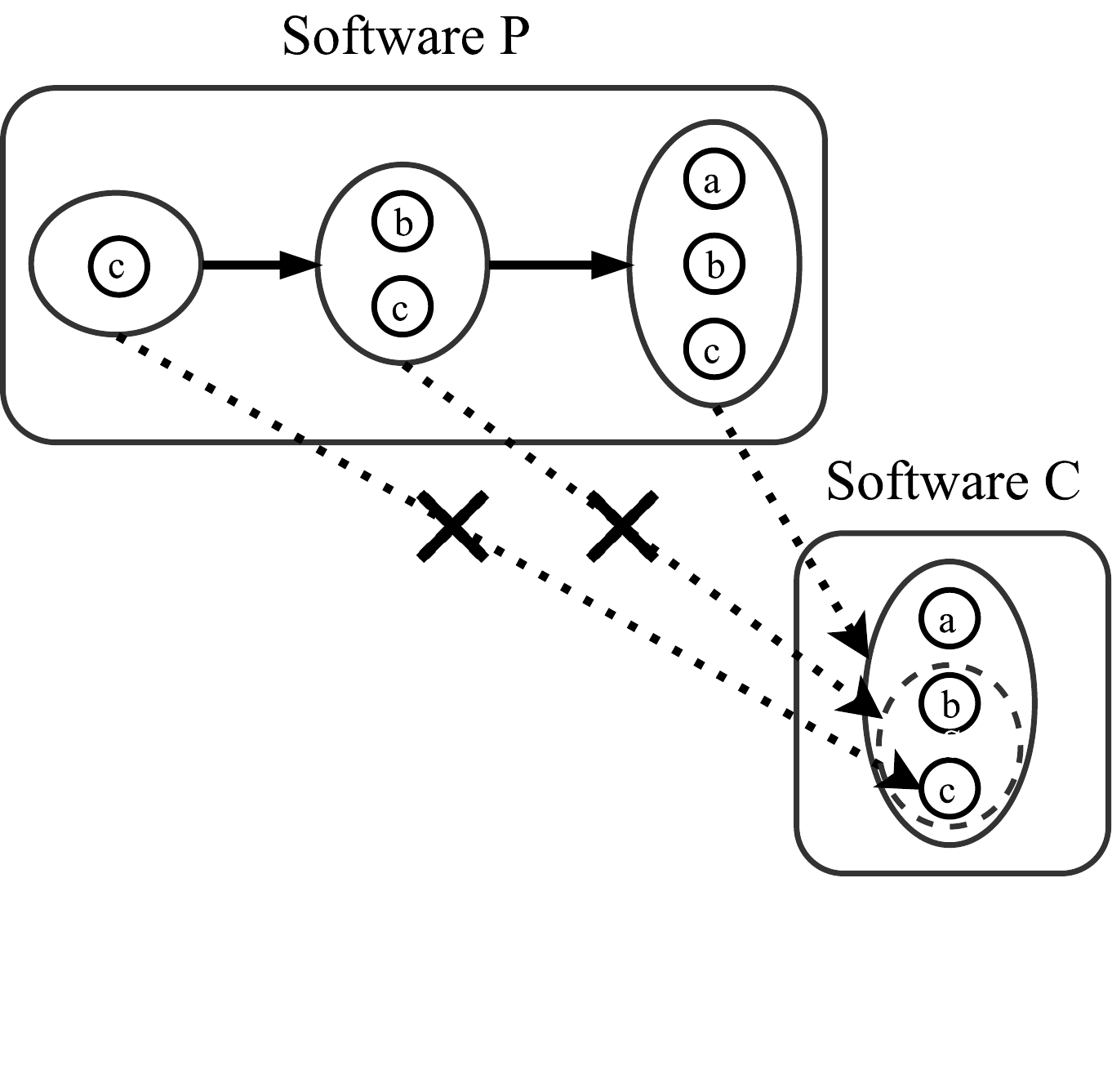} 
\end{tabular}
}%
\label{fig:c2}
\hfil
\subfloat[C3: most similar file sets]{
\begin{tabular}{c}
\includegraphics[width=0.15\linewidth]{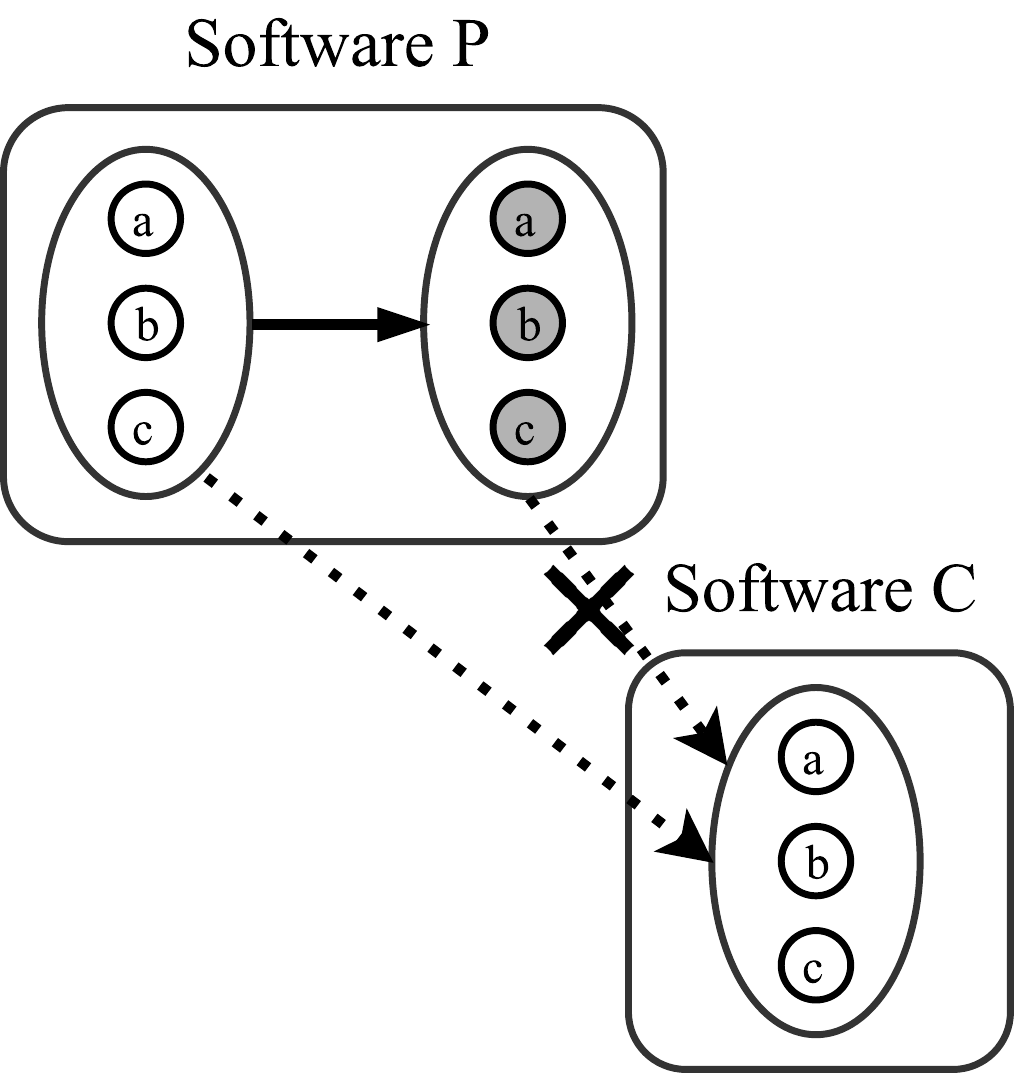} 
\end{tabular}
}%
\label{fig:c3}
\hfil
\subfloat[C4: oldest file sets]{
\begin{tabular}{c}
\includegraphics[width=0.15\linewidth]{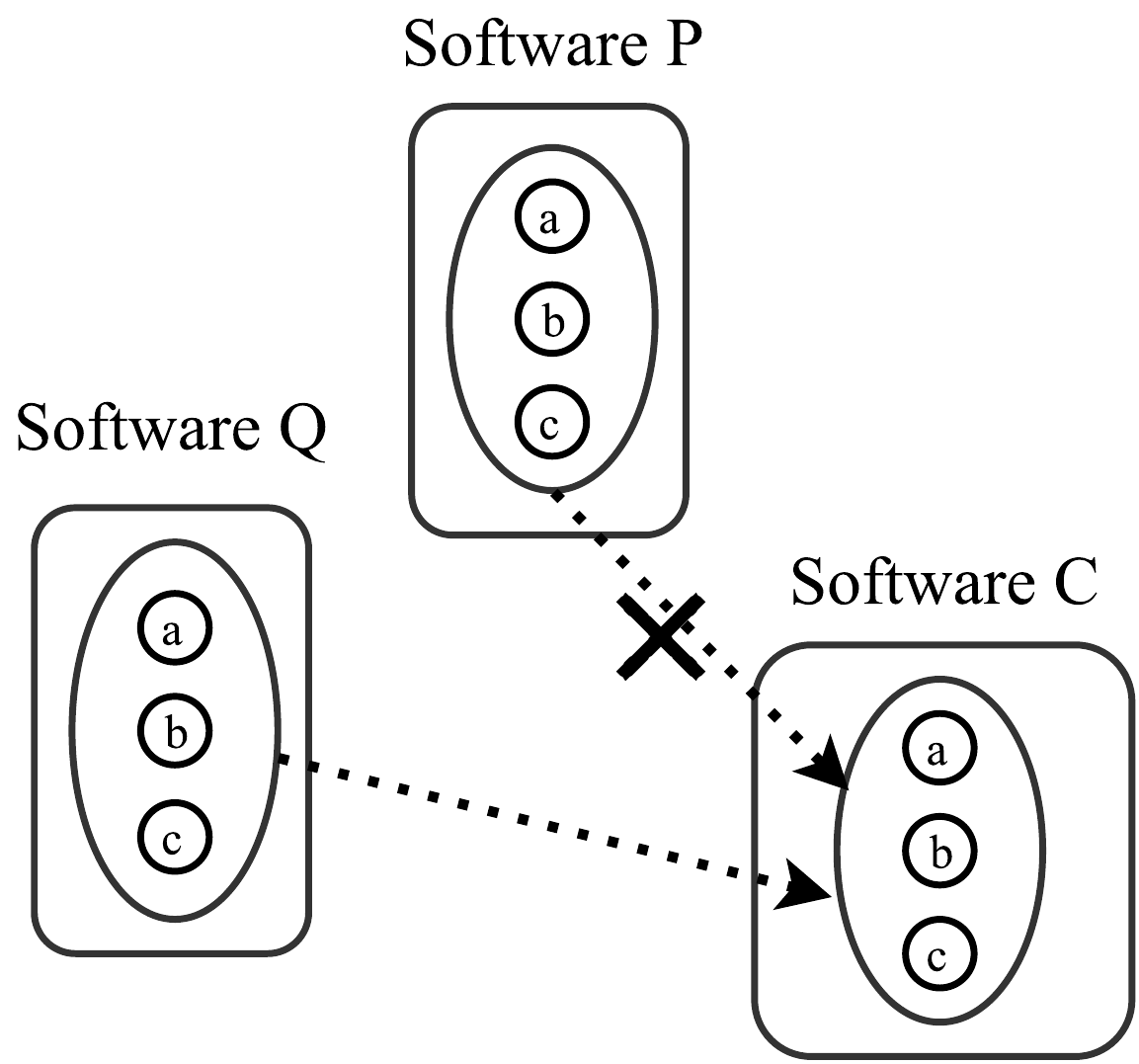} 
\end{tabular}
}%
\label{fig:c4}
\hfil
\subfloat[C5: single origins]{
\begin{tabular}{c}
\includegraphics[width=0.15\linewidth]{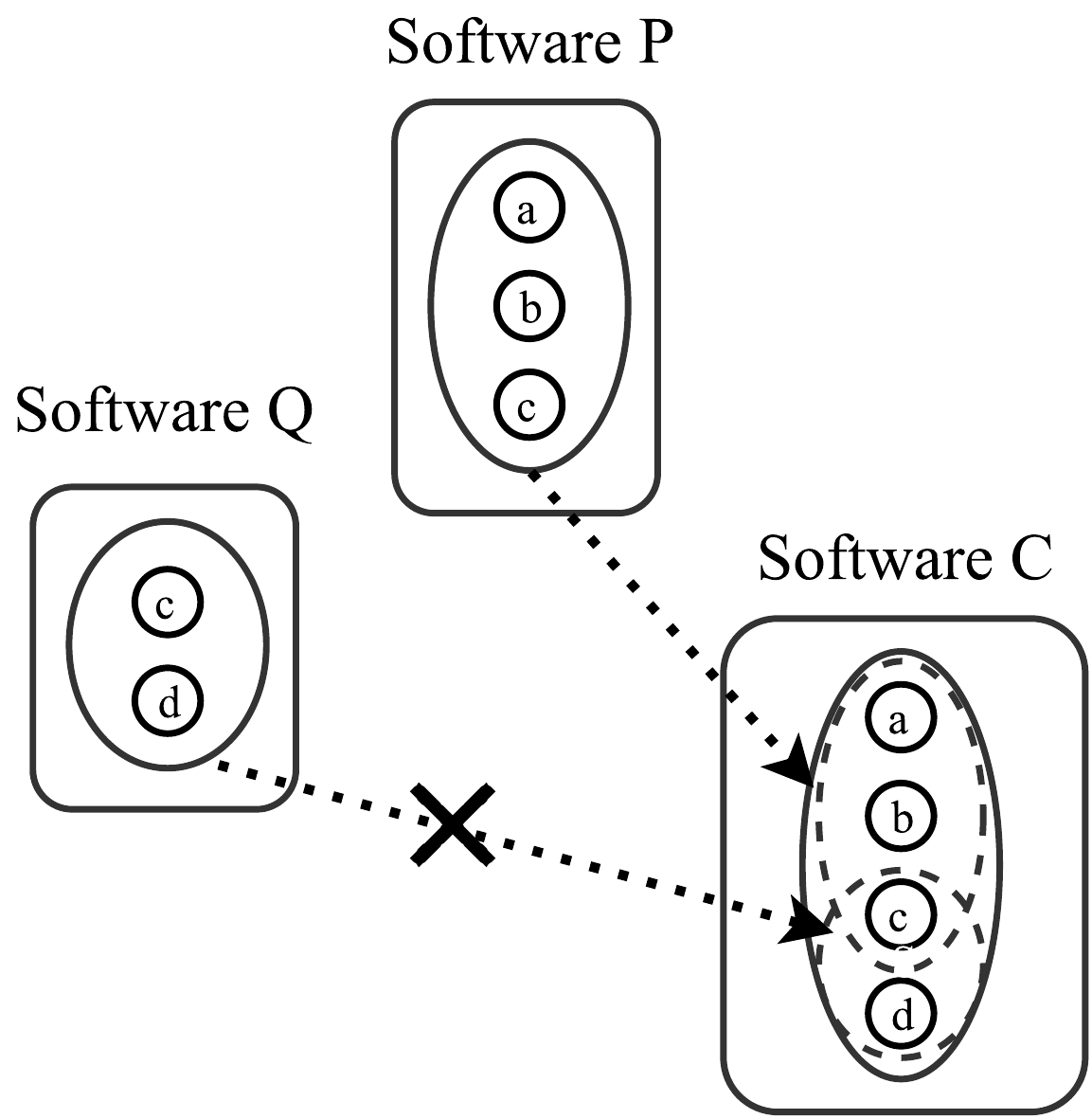} 
\end{tabular}
}%
\label{fig:c5}
\hfil
\subfloat[C6: update rather than clone-and-own]{
\begin{tabular}{c}
\includegraphics[width=0.15\linewidth]{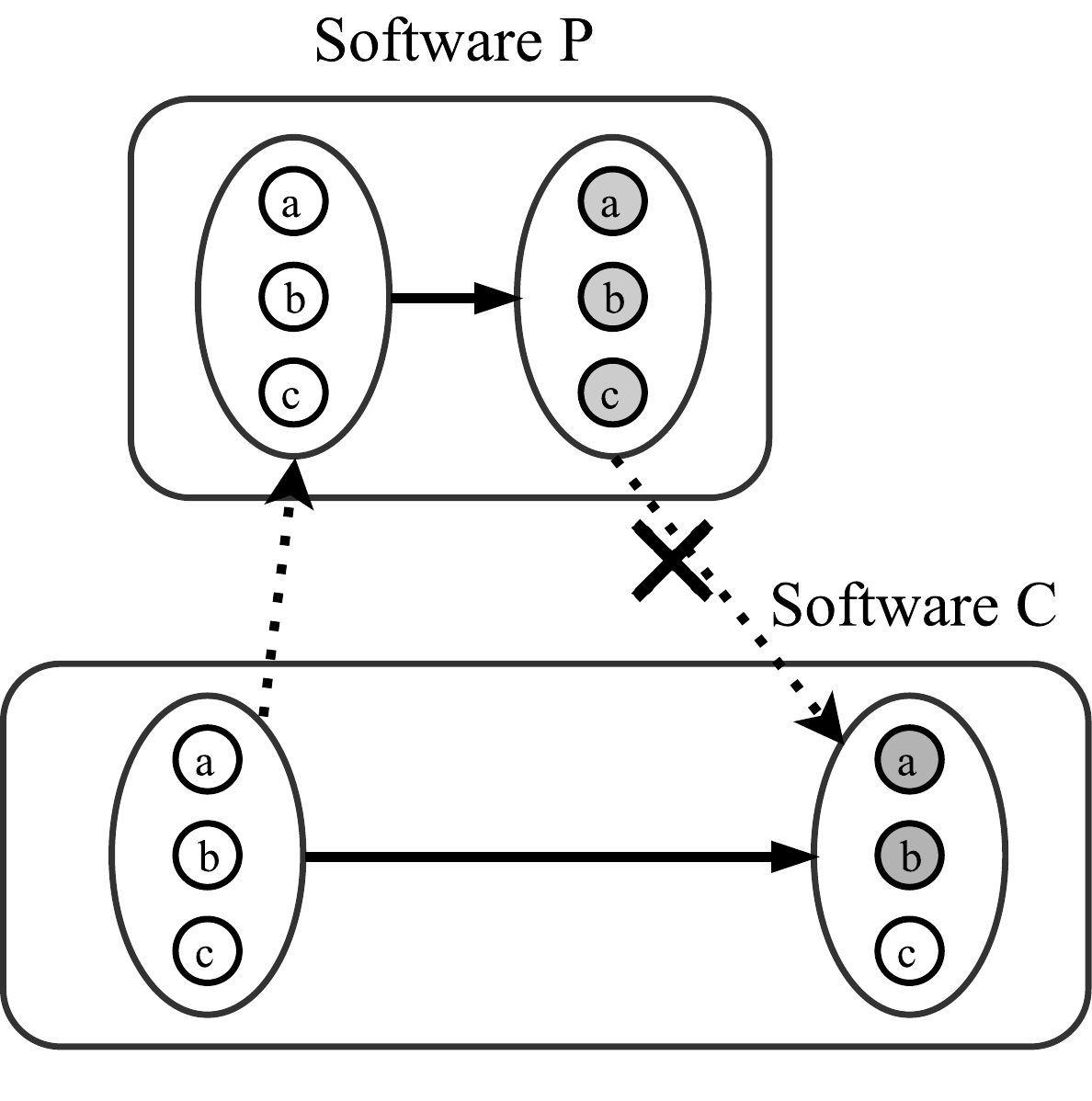} 
\end{tabular}
}%
\label{fig:c6}
\caption{Criteria to remove unlikely clone-and-own links. Solid lines represent update relations and dotted lines represent clone-and-own relations.}
\label{fig:criteria}
\end{figure*}

To identify clone-and-own operations among multiple projects, we link the original source file sets (\textbf{origins}) to the reused source file sets (\textbf{destinations}), in any combinations of source file set pairs from multiple software projects.
We call projects that have origins as \textbf{producer} projects, and reusing projects as \textbf{consumer} projects. 
In this study, we only target file sets belonging to tagged snapshots in Git repositories.
This is efficient to reduce the search spaces, and we think it reasonable to assume that developers reuse file sets in released snapshots, which should have tags.

A pair of file sets that have similar or identical files can be considered to be potential \textit{origin-destination} relations.
%
As reported in previous studies~\cite{Lopes:2017:DMC:3152284.3133908,Gharehyazie:2017:HCC:3104188.3104225},
there are a considerable amount of duplication in source code among multiple projects, which can be considered to be potential origin-destination relations.
We prepare criteria for choosing most likely clone-and-own relations (Section~\ref{sec:criteria}), and design an algorithm of detecting many-to-many clone-and-own relations (Section~\ref{sec:algorithm}).

\subsection{Criteria for Clone\&Own Detection}
\label{sec:criteria}

To find most likely origins, we set the following criteria.
Among various potential clone-and-own relations, we remove unlikely relations based on our criteria as shown in Figure~\ref{fig:criteria}.

\begin{description}
\item[C1] \textbf{Origins come from the past.}
For each snapshot in a \textit{potential} consumer project, origin file sets are searched from the past (until the commit time of the snapshot).

\item[C2] \textbf{The largest file sets are more likely to be origins.}
Although a destination file set can compose of multiple small origins in theory, it is not reasonable to assume that developers pick a single or a few files from many different projects, instead of fetching set of files from specific projects.
In software P, there are three snapshots that contain one, two, and three files, which can correspond to subsets of files in the snapshot of software C (Figure~\ref{fig:criteria} (a)). Since the number of corresponding files is the largest, the third snapshot is likely to be the origin.

\item[C3] \textbf{Most similar file sets are more likely to be origins.}
Identical or similar file sets can be considered to be clone-and-own relations. As shown in Figure~\ref{fig:criteria} (b), the file set in software C is more similar to the file set in the first snapshot than the second snapshot in software P. Then we can assume that the first snapshot is likely to be the origin.

\item[C4] \textbf{Oldest file sets are more likely to be origins.}
If there are multiple potential origins that are in the same conditions with the criteria C2 and C3, we consider the oldest file sets is the origin. Considering Figure~\ref{fig:criteria} (c), the snapshot in software Q is likely to be the origin.

\item[C5] \textbf{Each file does not come from multiple origins.}
There can be multiple potential origins for single files. In the example shown in Figure~\ref{fig:criteria} (d), the file $c$ is associated with both snapshots in software P and Q.
Based on the criteria C2, C3, and C4, only one origin is selected for a destination that have overlapped potential origins (the snapshot in software P is selected, in this example).

\item[C6] \textbf{Direct ancestors are more likely to be origins.}
If file sets can be considered to be updated within a project, potential clone-and-own links are removed. In Figure~\ref{fig:criteria} (e), the second snapshot in software P can be identified as a temporal origin for the second snapshot in software C, based on the above criteria. However, the previous version of the snapshot exists in software C. Moreover, files a, b, and c in software P originally come from software C. Then we consider that the second snapshots in both software P and C become occasionally similar, or, at least, there is no clone-and-own relation between them.
\end{description}

\subsection{Clone\&Own Detection Algorithm}
\label{sec:algorithm}

\begin{figure*}[!t]
\centering
\subfloat[Step 1: collect origin candidates]{
\begin{tabular}{c}
\includegraphics[width=0.25\linewidth]{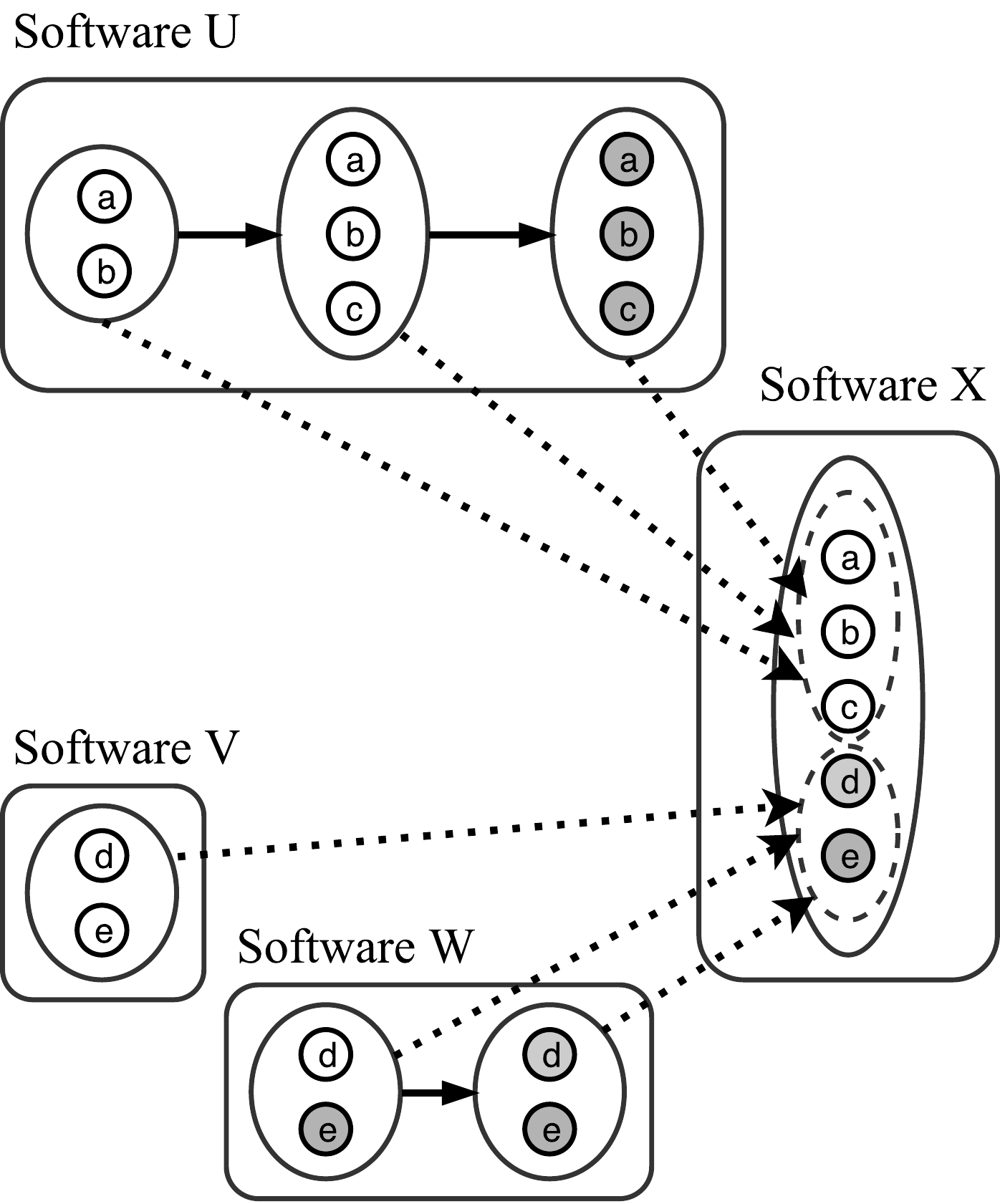} 
\end{tabular}
}%
\label{fig:s1}
\hfil
\subfloat[Step 2: select one origin for each project]{
\begin{tabular}{c}
\includegraphics[width=0.25\linewidth]{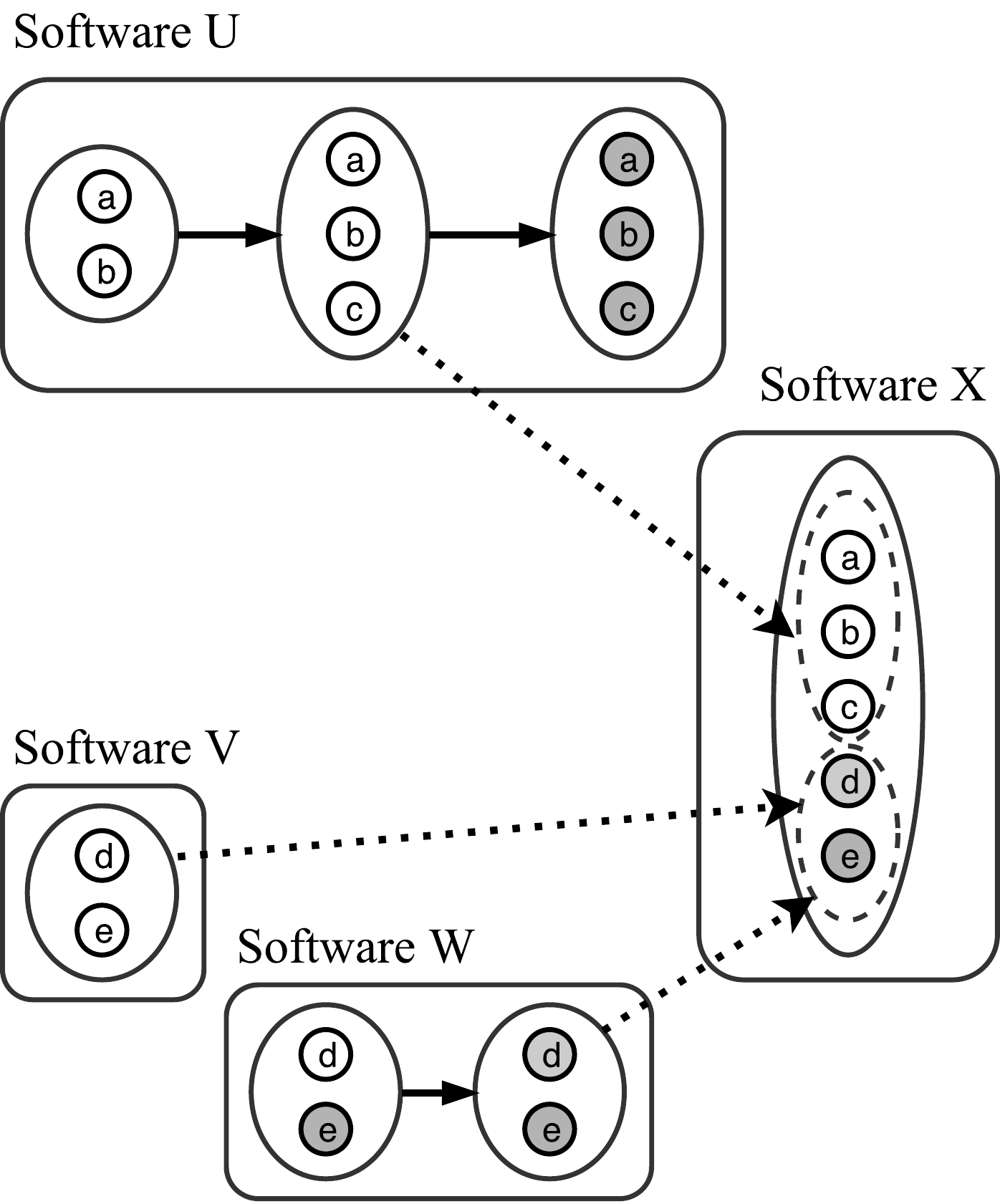} 
\end{tabular}
}%
\label{fig:s2}
\hfil
\subfloat[Step 3: determine temporal clone-and-own links]{
\begin{tabular}{c}
\includegraphics[width=0.25\linewidth]{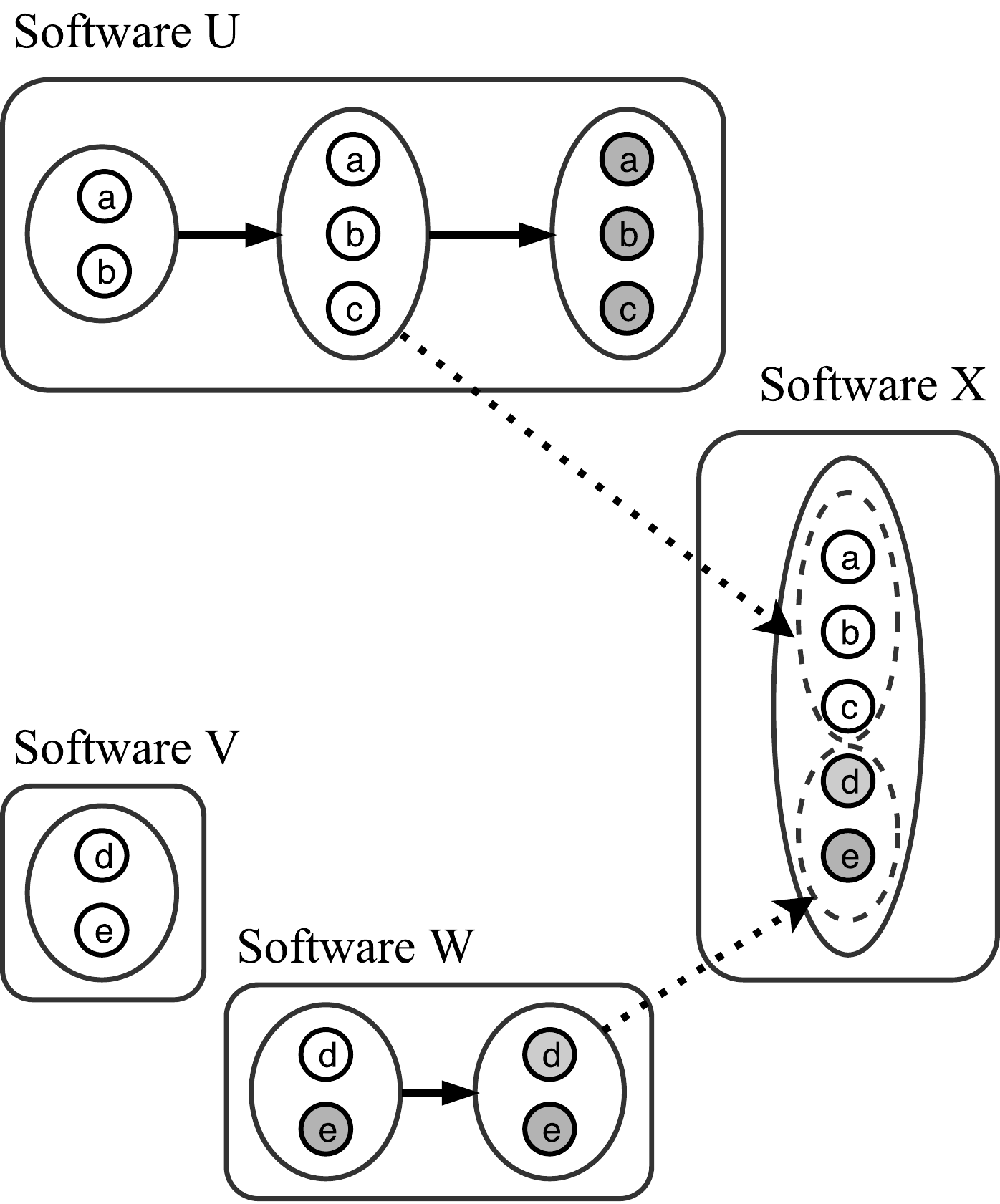} 
\end{tabular}
}%
\\
\label{fig:s3}
\subfloat[Step 4-I: remove links between identical files]{
\begin{tabular}{c}
\includegraphics[width=0.18\linewidth]{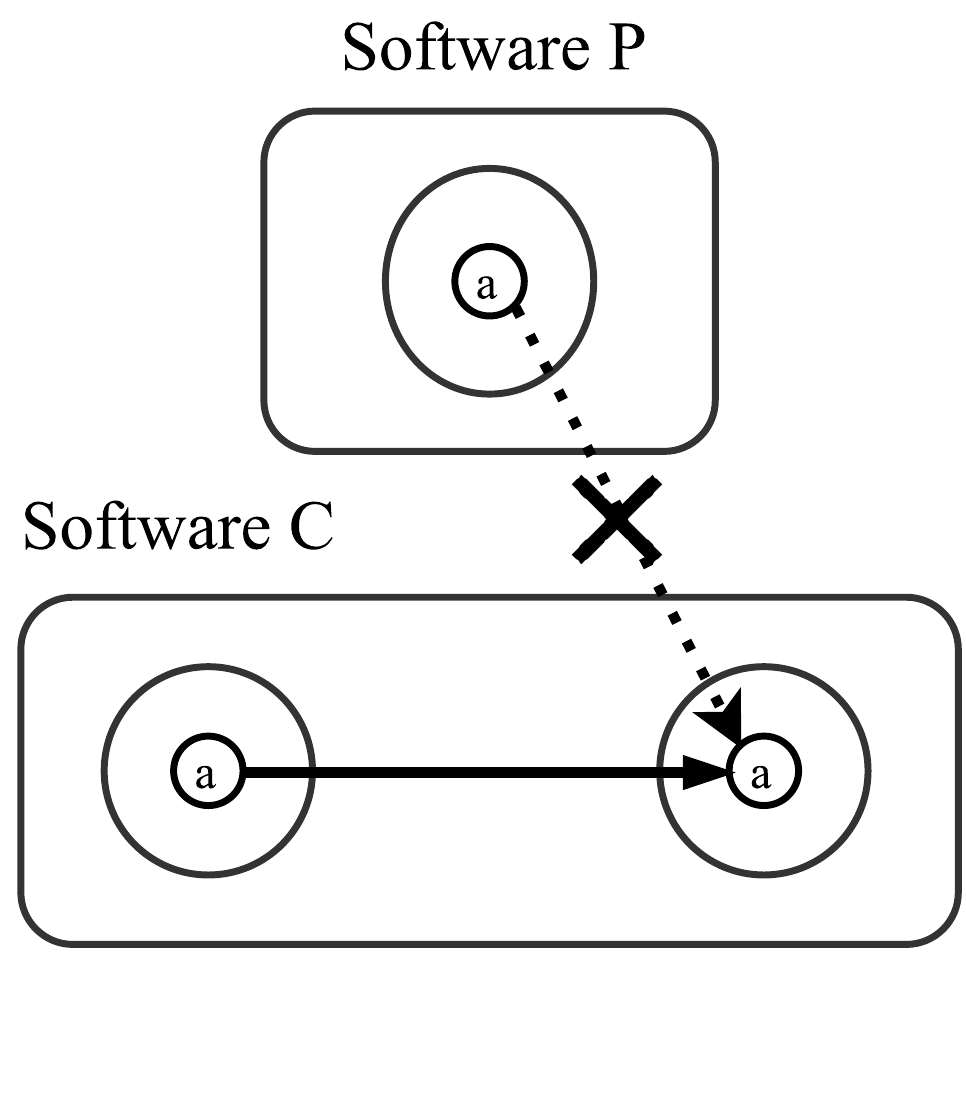} 
\end{tabular}
}%
\label{fig:s4a}
\hfil
\subfloat[Step 4-II: remove links between similar files]{
\begin{tabular}{c}
\includegraphics[width=0.18\linewidth]{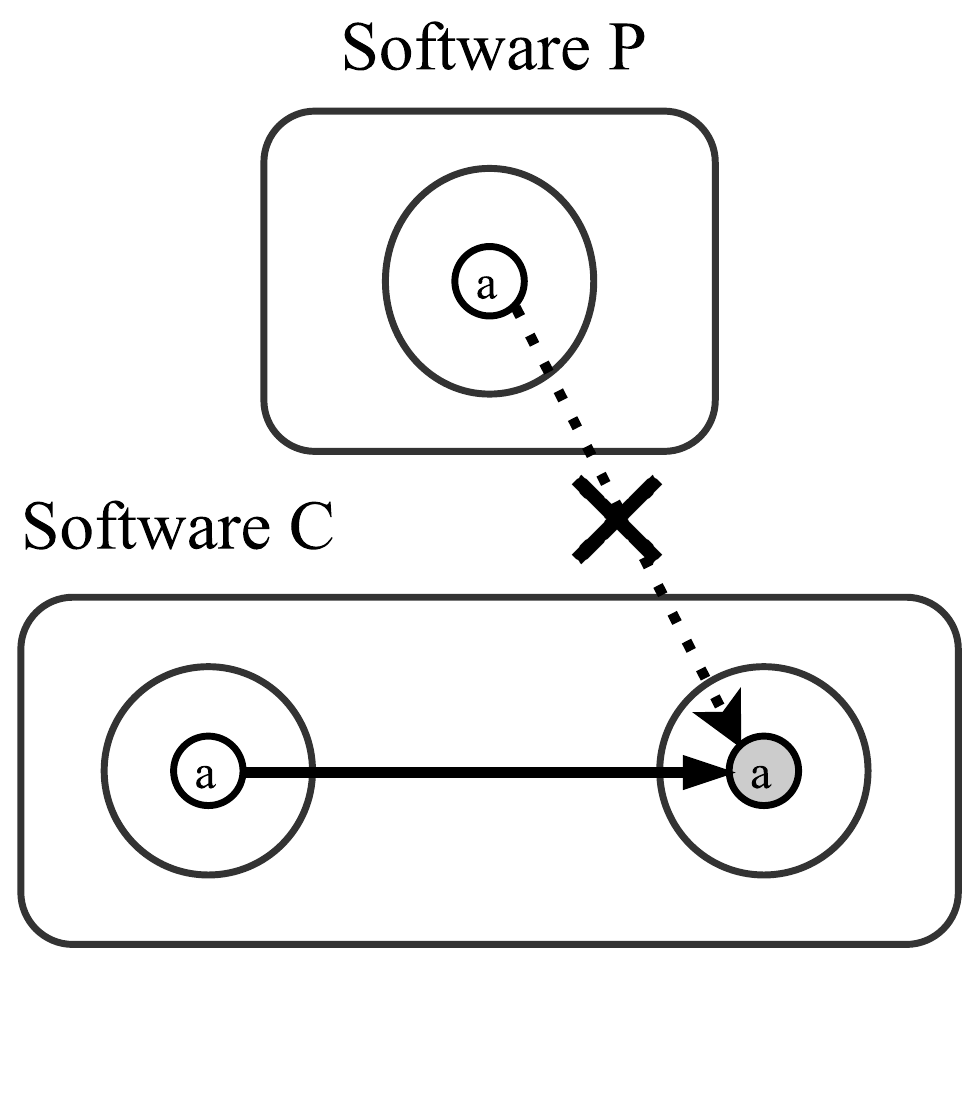} 
\end{tabular}
}%
\label{fig:s4b}
\hfil
\subfloat[Step 4-III: remove false transferring links]{
\begin{tabular}{c}
\includegraphics[width=0.18\linewidth]{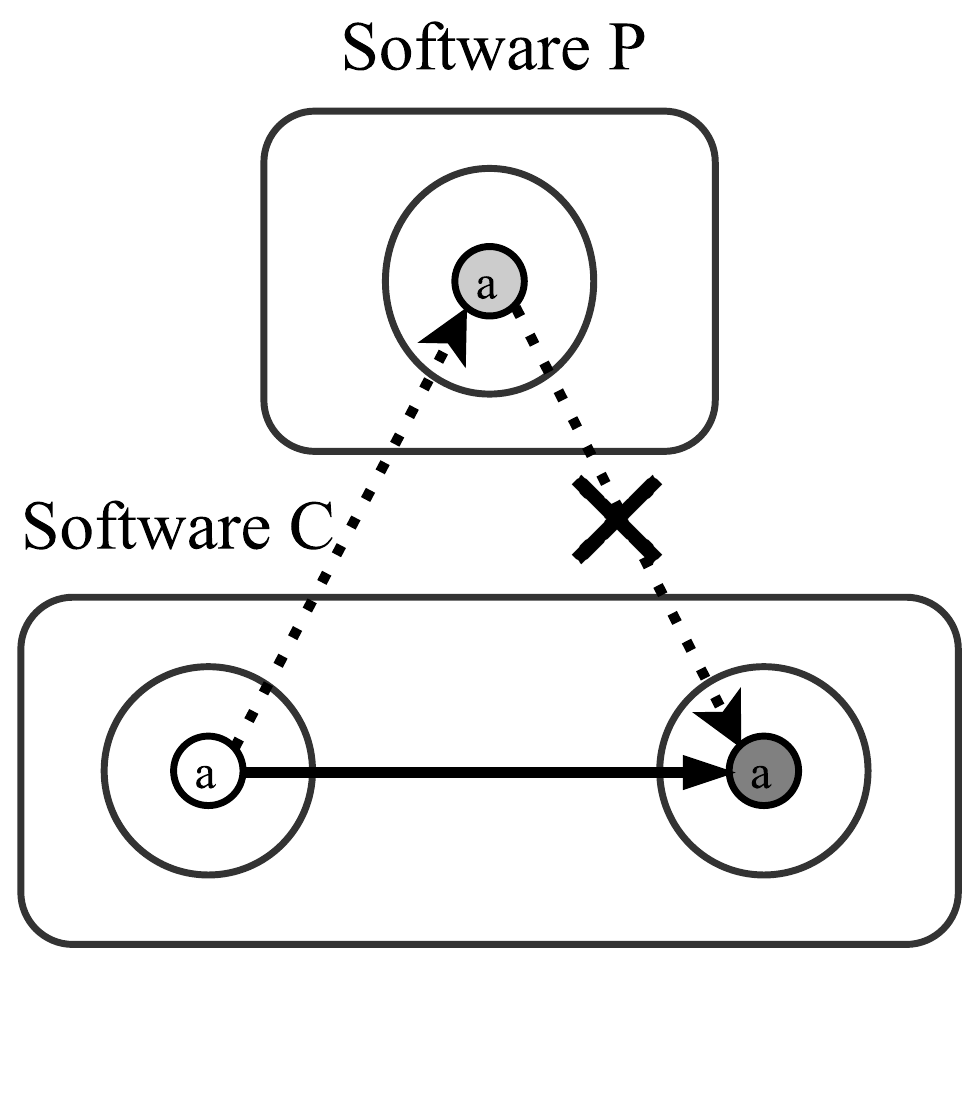} 
\end{tabular}
}%
\label{fig:s4c}
\caption{Clone-and-own detection algorithm. Solid lines represent modifications within projects and dotted lines represent clone-and-own relations.}
\label{fig:step1234}
\end{figure*}

Given a list of Git repositories of software projects, we construct software supply chain maps among the projects.
Our algorithm consists of five steps.
For all snapshots in the projects, potential origin file sets are searched by performing the following three steps. Figure~\ref{fig:step1234} (a) to (c) illustrate these steps with an example of the snapshot in Software X as an query.
Since we do not know actual developers' intentions of clone-and-own operations in the past, we try to identify only plausible operations and ignore suspicious relations.

\begin{description}
\item[Step 1] \textbf{Collect candidates of origins.}
For every file in the query snapshot, identical or similar files are searched from any snapshots in other projects, which had been created before the query snapshot (\textbf{C1} criterion).
We identify identical files using SHA-1 file hash. 
We also identify similar files using n-gram-based similarity defined in \cite{Ishio:2017:SFS:3104188.3104222} because of its efficiency for our large-scale analysis.
Found files in a snapshot are aggregated as one file set.
If there is no candidate origin for a query snapshot, terminate the origin detection for the snapshot.
In the example shown in Figure~\ref{fig:step1234} (a), identical or similar files are found in five different snapshots from three projects.

\item[Step 2] \textbf{Select one origin candidate for each potential producer project.}
From multiple origin candidates in one potential producer project, we decide a winner candidate. Multiple origin candidates are sorted based on the following conditions (if there is a tie, the next condition is applied to the tie).
The first candidate in the sorted list is chosen as a winner.
\begin{enumerate}
\item The descending order of the number of files in file sets (\textbf{C2} criterion).
\item The descending order of the number of identical files in file sets (\textbf{C3} criterion).
\item The ascending order of the committed times of file sets (\textbf{C4} criterion).
\end{enumerate}
In the example shown in Figure~\ref{fig:step1234} (b), links from lost origin candidates are removed.

\item[Step 3] \textbf{Determine temporal clone-and-own relations.}
From multiple potential producer projects, which have single winner origins, we decide winner projects if there are overlapped origins (\textbf{C5} criterion).
Among overlapped origins, only one origin is selected based on the same sorting approach with step 2.
In the example shown in Figure~\ref{fig:step1234} (c), software V is removed and the remained links are considered to be temporal clone-and-own relations.
\end{description}

In the temporal clone-and-own relations obtained from the above three steps, there can be inappropriate detection. To remove inappropriate clone-and-own relations, the following steps are performed.

\begin{description}
\item[Step 4] \textbf{Remove files that come from own projects.}
In this step we investigate the history of each file in temporal destination file sets. If files come from their own repositories, it is inappropriate to consider that they were fetched from other repositories (\textbf{C6} criterion). We remove temporal clone-and-own relations between files if their relations are one of the following three types shown in Figure~\ref{fig:step1234} (d) to (f). In the figures, software P is considered to be a temporal producer, and software C is a temporal consumer.
\begin{enumerate}
\item Linked two files are identical, and the file exist in a temporal consumer project before a temporal origin is created (Figure~\ref{fig:step1234} (d)).
\item Linked two files are similar, and the temporal origin file exists in a temporal consumer project (Figure~\ref{fig:step1234} (e)).
\item Linked two files are similar, and the temporal origin file does not exist in a temporal consumer project. However, the temporal origin file come from a temporal consumer project (Figure~\ref{fig:step1234} (f)).
\end{enumerate}

\item[Step 5] \textbf{Remove suspicious clone-and-own relations.}
We prepare two rules to remove suspicious relations and leave more probable relations.
\begin{enumerate}
\item Remove a clone-and-own link from software B to A if there is an older link from software A to B.  We assume that clone-and-own reuses happen one-way and returning reuses rarely happen.
\item Remove a clone-and-own link whose corresponding file set size (the number of files) is less than a threshold. This is because a small number of files can occasionally become similar to files in other projects. Threshold values are discussed in Section~\ref{sec:process}.
\end{enumerate}
\end{description}

By identifying producers and consumers (then identifying `reuse' operations), we can track the histories of files across projects, detect the transitions of ownership, and consider file versions across projects. As far as we know, this is the first study of identifying producers and consumers in file set reuses (clone-and-own operations) and showing chain maps (file set reuse relations).

\subsection{Algorithm Validation}

Our algorithm is validated with the manually labeled clone-and-own relations in \texttt{Firefox 45.0} and \texttt{Android 4.4.2\_rc1} used in the previous study~\cite{Ishio:2017:SFS:3104188.3104222}.
We identified appropriate origins in all destinations (21 from Firefox and 54 from Android).
Incorrect 14 false links (top-1 in the ranking) in the previous study were successfully excluded and correct destinations are found because of the time consideration in our algorithm.
With this encouraging result, we conduct an empirical study of software supply chain maps in FLOSS projects.

\section{Study Procedure}

We study software supply chain maps in FLOSS projects with the following three research questions.

\RqOne~ We analyze network metrics to discuss the characteristics of obtained software supply chain maps.

\RqTwo~ We investigate how producer projects evolved along with the growth of software ecosystems.

\RqThree~ The nature of supply chains in software supply chain maps are analyzed.

To study clone-and-own relations among actual projects, we try to construct chain maps only with plausible links by ignoring false positive relations. To this end, we prepare relatively strict settings of threshold to detect clone-and-own relations conservatively.

\subsection{Project Collection}

We collected software projects written in C, C++, or Java from GitHub.
For each language, we selected highly starred projects as they considered to be influential.
In order to include both applications and libraries, we used three search queries for each language.  
Queries ``library'' and ``lib'' were used to identify library projects.
A query ``mirror'' was used to select popular projects developed outside of GitHub but having a mirror repository in GitHub.
For each query, we obtained up to 1,000 projects that have at least 20 stars.
In addition, we also collected up to 1,000 projects without query keyword under the same star condition, i.e. the most popular projects for each programming language.
The dataset includes 7,468 projects.
Their Git repositories were cloned on November 18, 2017.  
The total size is 341 GB.

\subsection{Data Processing}
\label{sec:process}

We analyzed tagged snapshots (e.g. official releases) of files in the Git repositories.
%
The grammars of C++ 14 and Java 8 were used for our lexical analysis.
The lexical analysis tools excluded whitespace and comments from tokens, but keeps preprocessor directives for C/C++.
We extracted trigrams from each of unique files, performed pair-wise comparison, and then computed file similarities across projects.

\begin{table}
\caption{Stepwise clone-and-own detection}
\label{tab:step}
\begin{tabular}{crrr}
\toprule
Step & \# of projects & \# of links & \# of associated files \\
\midrule
0 & 4,592 & --- & (9,745,519) \\
2 & 4,592 & 111,516 & 541,795 \\
3 & 2,330 & 5,808 & 391,946 \\
4 & 1,847 & 3,306 & 126,689 \\
\bottomrule
\end{tabular}
\end{table}

From the collected projects, we removed projects that did not have tagged snapshots containing targeted source files. The number of left projects was 4,592 and there were 9,745,519 unique files in the entire tagged snapshots in the projects, as shown in the step 0 row of Table~\ref{tab:step}.
We can see, from Table~\ref{tab:step}, how our algorithm constructs most likely clone-and-own links by removing inappropriate links step by step.
After step 2, all projects belonged to the temporal network with a large amount of links, which means that those analyzed projects have at least one identical or similar file with other projects.
After step 3, unlikely clone-and-own links were largely removed, and false producer and consumer projects were removed from the temporal network.
After step 4, more than two-thirds of the associated files, which are considered to be inappropriate, were removed, and then the number of projects and links decrease. 
With the first rule in step 5, about 100 links are removed (the number of links became 3,203).

\begin{figure}
\includegraphics[width=.9\linewidth]{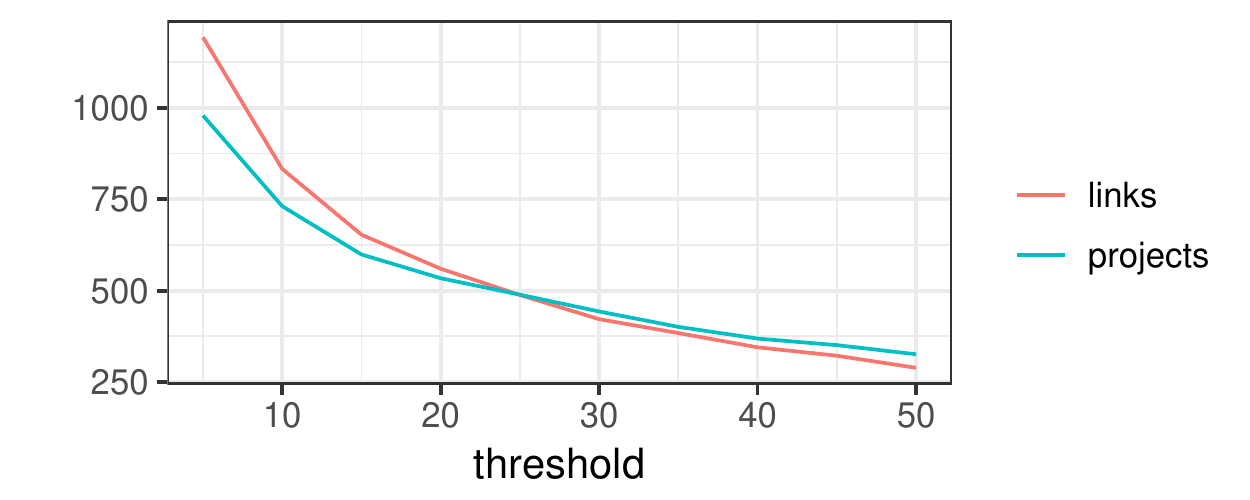}
\caption{Step 5-2: removing small file set size links.}
\label{fig:threshold}
\end{figure}

\begin{figure}
\includegraphics[width=.9\linewidth]{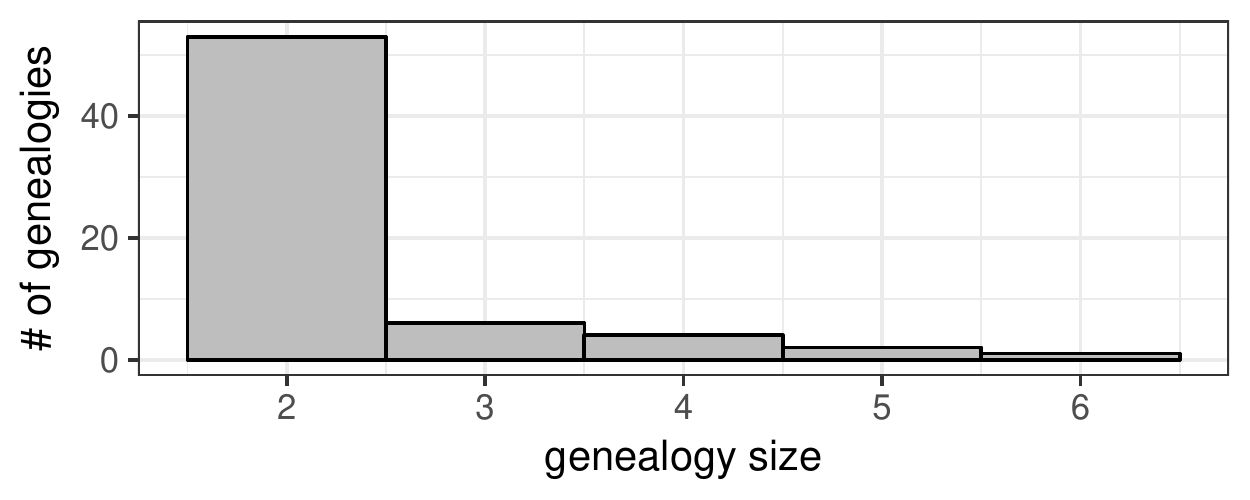}
\caption{Distributions of the constructed supply chain maps. The largest map (size 383) is excluded.}
\label{fig:size}
\end{figure}

Figure~\ref{fig:threshold} illustrates the impact of the threshold in the second rule of step 5. The threshold changes from 5 to 50.
We find the about half of the projects are involved in less than five file relations (from 1,847 projects without threshold to less than 1,000 at the threshold 5). Although there can be small clone-and-own reuses, there also can be occasional matches. In order to focus on intentional clone-and-own reuses, we use the threshold value 20 in this study.
This threshold will reduce smaller repositories; Lopes et al. reported 9 for Java and 11 for C++ as the median value of the number of files in a projects in their large-scale study~\cite{Lopes:2017:DMC:3152284.3133908}.

With the above settings, 534 projects are identified as members of software supply chain maps. There are 67 independent maps, from small size maps only with two belonging projects (size 2) to the largest map (size 383). Figure~\ref{fig:size} presents the histogram of the map sizes excluding the largest one. Although there are many small maps, more than $70\%$ (383/535) of projects belong to one big map shown in Figure~\ref{fig:network}. The obtained map data is now available\footnote{\url{https://www.dropbox.com/s/q25w60ixu7xu1a6/genealogy.zip?dl=0}}.
We plan to prepare archived open data in case of acceptance.

\begin{figure}
\includegraphics[width=.9\linewidth]{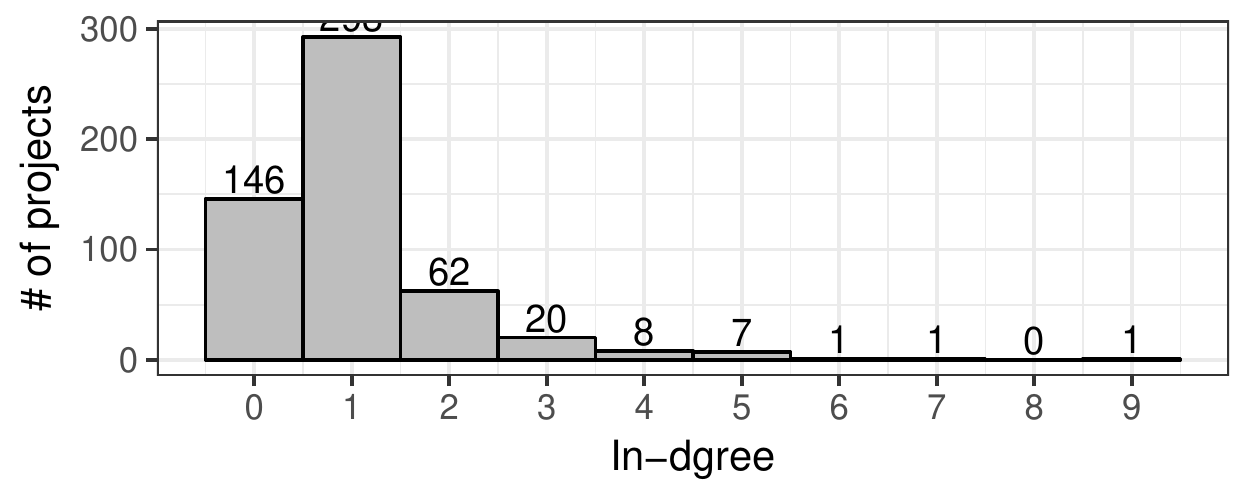}
\caption{In-degree (reusing) distributions.}
\label{fig:indegree}
\end{figure}

\begin{figure}
\includegraphics[width=.9\linewidth]{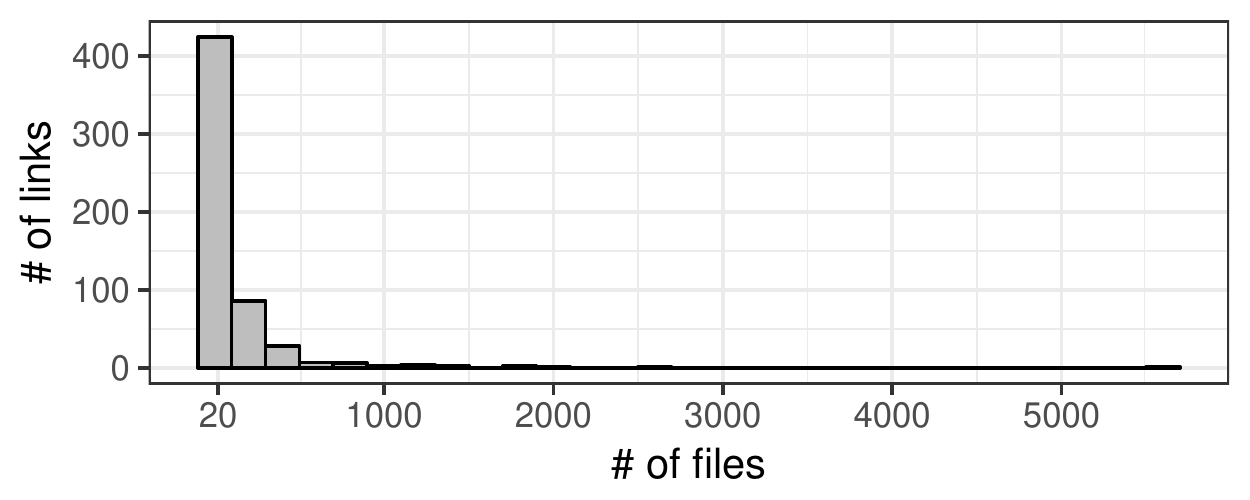}
\caption{Distributions of the reused file set sizes.}
\label{fig:files}
\end{figure}

\begin{figure}
\includegraphics[width=.9\linewidth]{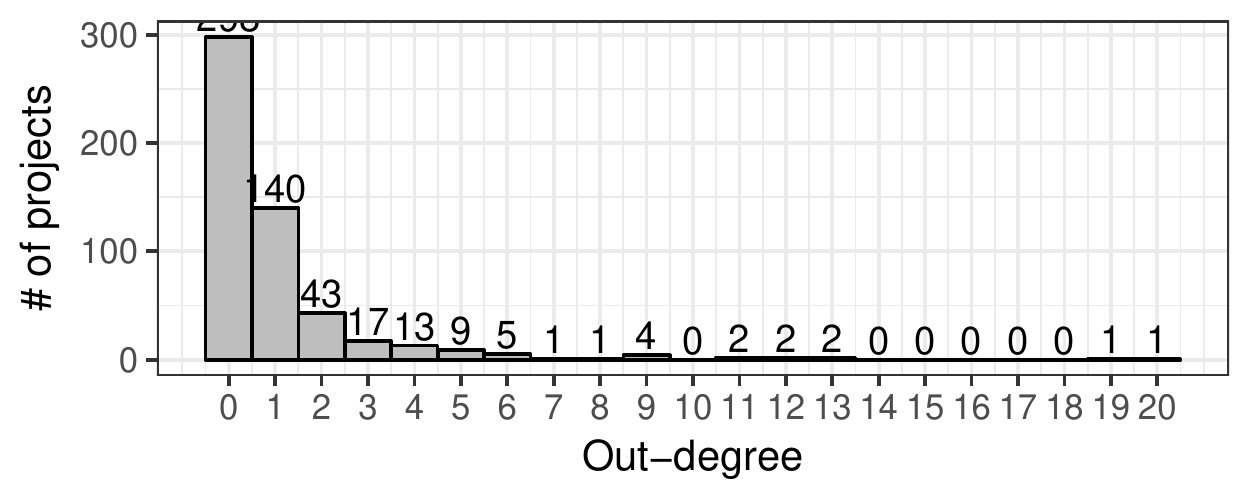}
\caption{Out-degree (reused) distributions.}
\label{fig:outdegree}
\end{figure}

\section{Study Results}

\subsection{RQ1: Clone\&Own Characteristics}

Now that we have seen that more than $10\%$ of analyzed projects belong to software supply chain maps, we proceed to analyze their software supply chain maps to reveal their characteristics.

Figure~\ref{fig:indegree} shows the distributions of reusing producer projects (in-degree) per consumer project. More than a quarter of projects belong to the maps only as producers, and most projects clone source files from single projects.
The distributions of the reused file set sizes can be seen in Figure~\ref{fig:files}.
More than three quarters of clone-and-own reuses (441 in the entire 560 links) 20 to 100 files.
There are 14 reuses contain more than 1,000 files from single producer projects, and the largest one involves 5,620 files.

Figure~\ref{fig:outdegree} presents the distributions of reused projects (out-degree) per producer project. We find that more than half of projects (298/534) were not reused by other projects, and most of the other projects were reused from only a few projects.
There are a small number of projects that are reused by more than 10 projects, and this finding is similar to the result in the previous work~\cite{Gharehyazie:2017:HCC:3104188.3104225}, that is, there are hubs of reusable code providers to other projects. We consider that the existence of clone-and-own hubs results in the similar observations in fine-grained clone detection work.

\begin{tcolorbox}
\textbf{Summary}: More than a quarter of projects played only producer roles, and more than half were just consumers.
In most clone-and-own cases, less than 100 files were reused, and only a few cases reused more than 1,000 files. There are a small number of projects that serves as hubs of source file set providers.
\end{tcolorbox}

\subsection{RQ2: Supply Chain Map Evolution}

\begin{figure}
\includegraphics[width=.8\linewidth]{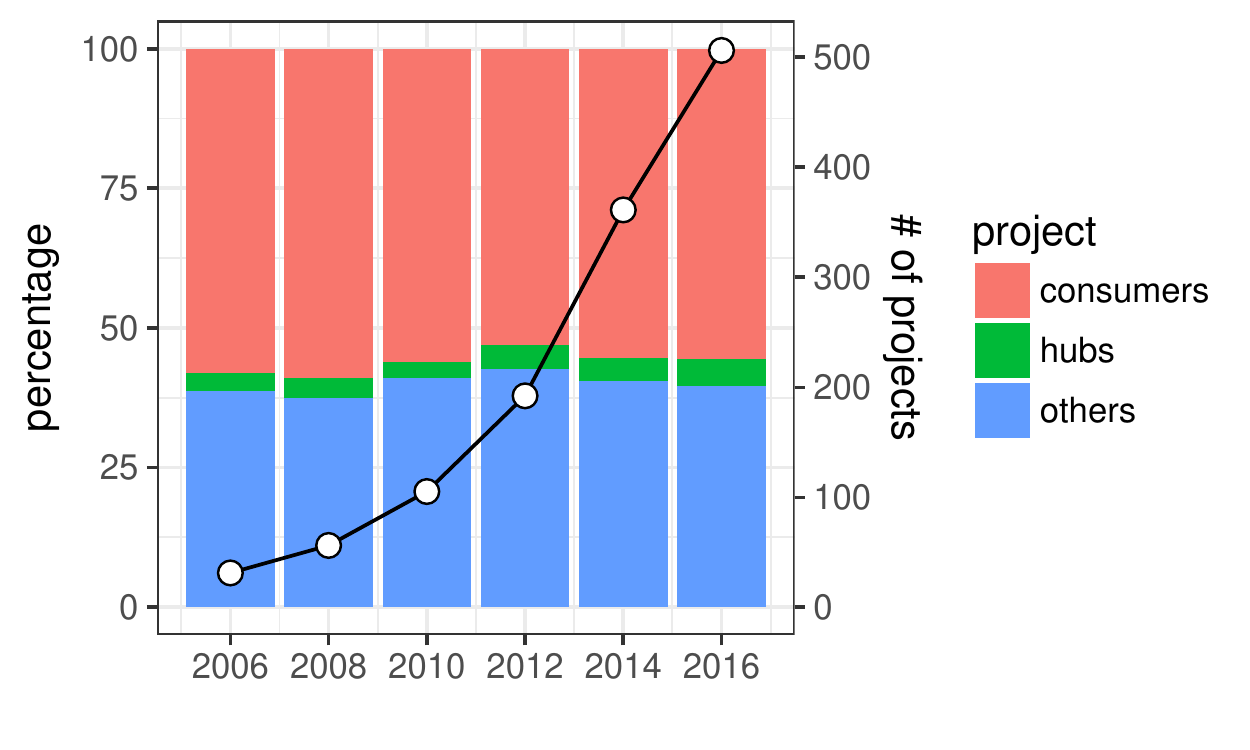}
\caption{Growth of the map sizes (a line) and the percentages of consumers, hubs, and the other projects (stacked bars).}
\label{fig:ecosystem}
\end{figure}

\begin{figure}
\includegraphics[width=.8\linewidth]{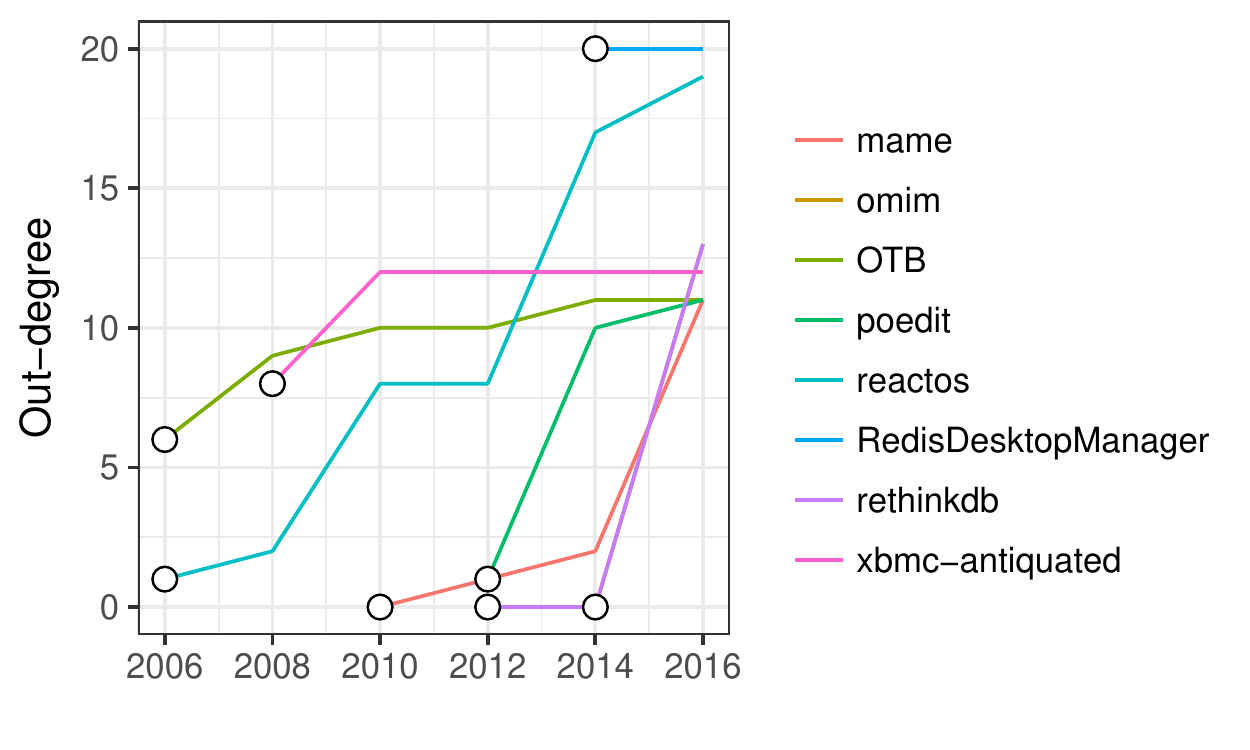}
\caption{Out-degree transitions of hub projects.}
\label{fig:hub}
\end{figure}

\begin{figure*}[!t]
\centering
\subfloat[2010]{
\begin{tabular}{c}
\includegraphics[width=0.3\linewidth]{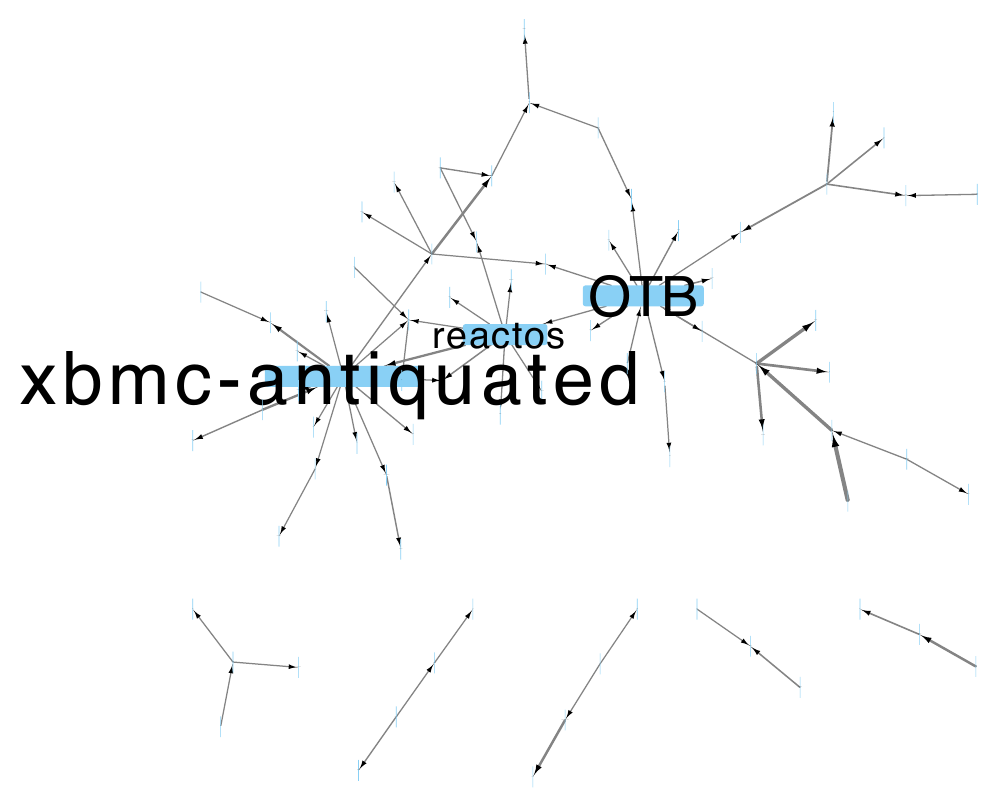} 
\end{tabular}
}%
\hfil
\subfloat[2014 (the largest map)]{
\begin{tabular}{c}
\includegraphics[width=0.3\linewidth]{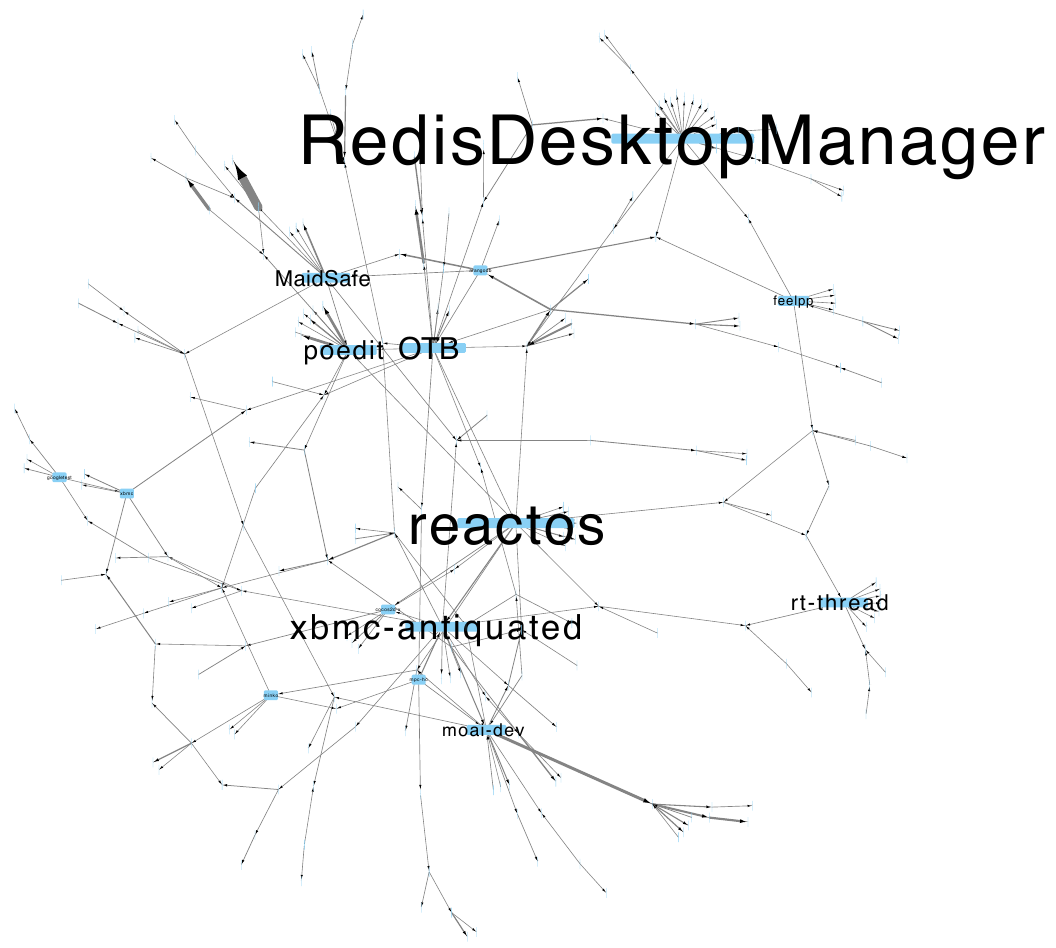} 
\end{tabular}
}%
\hfil
\subfloat[2016 (the largest map)]{
\begin{tabular}{c}
\includegraphics[width=0.3\linewidth]{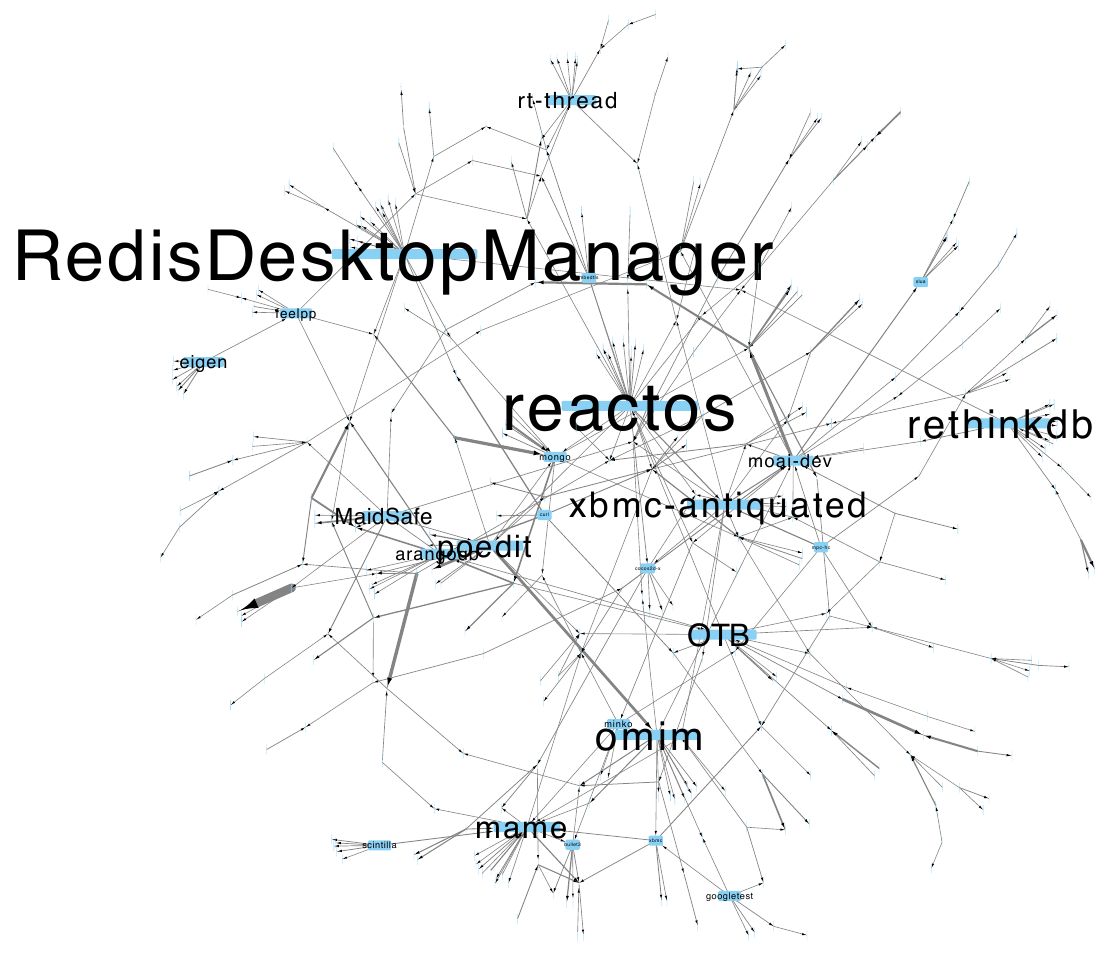} 
\end{tabular}
}%
\caption{software reuse map growth. Projects whose out-degrees are less than five are not shown.}
\label{fig:evol}
\end{figure*}

\begin{figure*}
\includegraphics[width=.9\linewidth]{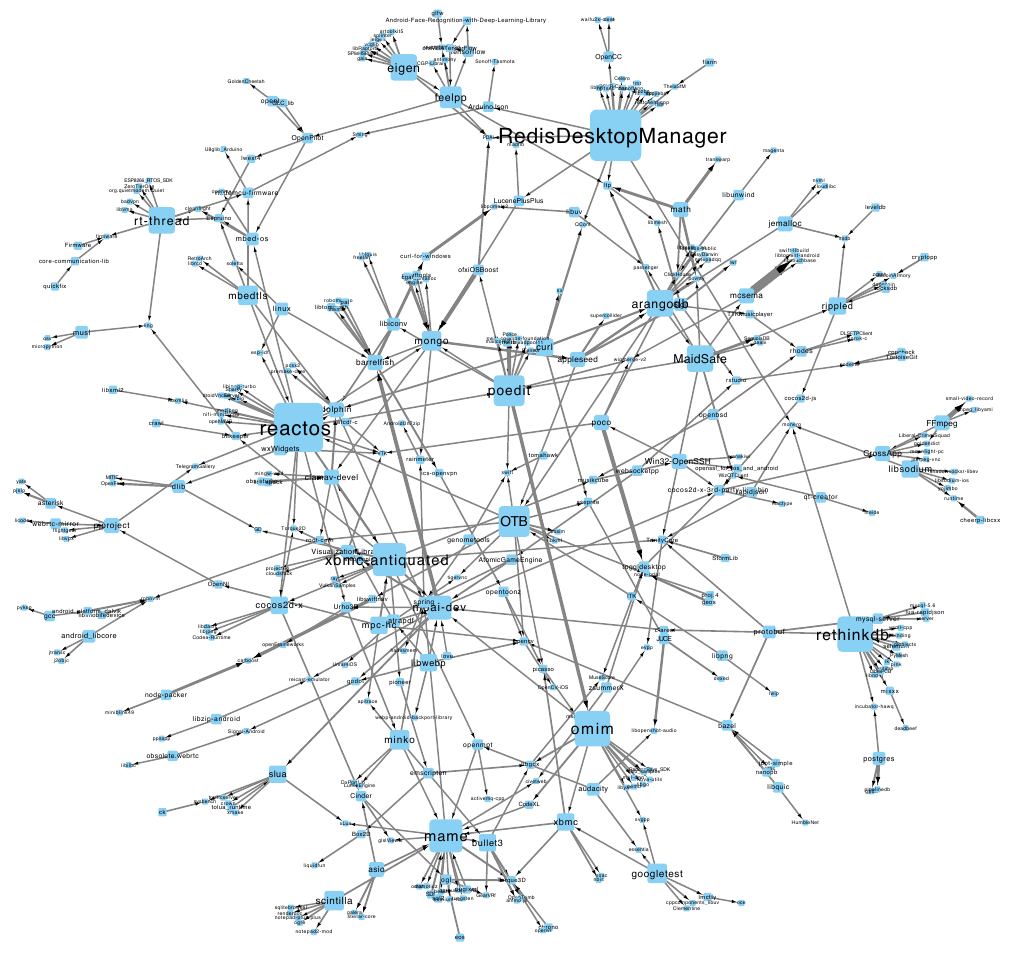}
\caption{The largest software supply chain map including 383 projects. The sizes of node represent the number of out-degree links and the width of links represent the number of reused files (in the first reuse).}
\label{fig:network}
\end{figure*}

\begin{figure}
\includegraphics[width=.9\linewidth]{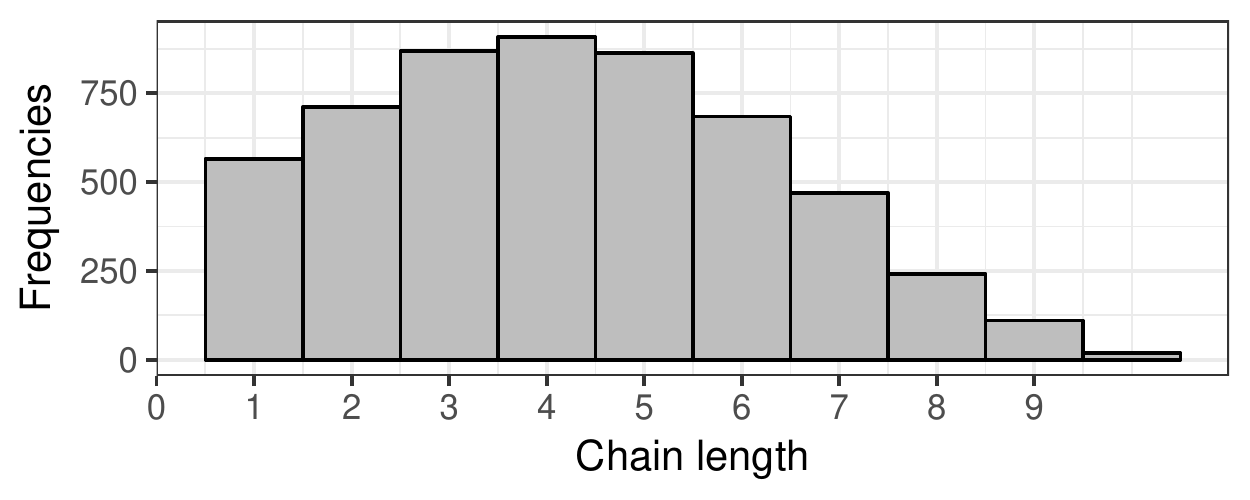}
\caption{Distributions of chain lengths among projects.}
\label{fig:chain}
\end{figure}

Figure~\ref{fig:ecosystem} shows how clone-and-own maps evolved from 2006 to 2016. A line illustrates the increase of the maps' members (projects), and stacked bars represent the percentages of consumer projects (no out-degree), hub projects (more than five out-degrees, which serves top 5\% projects), and the other projects.
Although the sizes of software supply chain maps largely increased from 2006 to 2016, their compositional structures are stable, that is, more than half of projects are consumers and there exist less than five percent hub projects. It can be said that inspired by a small number of hub software projects, newer projects are created by reusing them, and which makes software supply chain maps larger.

Next we focus on the evolution of hub projects.
As seen in Figure~\ref{fig:outdegree}, there are eight projects that have more than 10 out-degree links in the latest map.
Figure~\ref{fig:hub} shows the out-degree transitions of those eight projects.
We found two patterns of becoming hub projects: (i) being hubs from birth, and (ii) becoming hubs by connecting to the network. Projects \texttt{OTM}, \texttt{xbmc-antiquated}, and \texttt{RedisDesktopManager} belong to (i). These projects were widely reused from their early periods, regardless of week connections to supply chain maps. The out-degrees of the other projects in their early periods are relatively low. Project \texttt{reactos} increased its out-degree gradually after project \texttt{xbmc-antiquated} reused it in 2010 (Figure~\ref{fig:evol} (a)).
Similarly, project \texttt{poedit} increased its out-dgree when it connects to project \texttt{reactos} by using its source files in 2014 (Figure~\ref{fig:evol} (b)).
Project \texttt{mame} also increased its out-dgree when it reused source files of project \texttt{reactos} in 2016 (Figure~\ref{fig:evol} (c)). After connecting to the largest map in 2016, project \texttt{rethinkdb} was reused widely (Figure~\ref{fig:evol} (c)).
The final connections can be seen in Figure~\ref{fig:network}.

\begin{tcolorbox}
\textbf{Summary}: While the size of maps growing from 31 to 506 since 2006 to 2016, the percentages of consumer projects (just reusing others), hub projects (reused by more than five projects), and the other projects are stable: about $55\%$, less than $5\%$, and about $40\%$ respectively. Low out-degree projects can become hubs by connecting to hub projects.
\end{tcolorbox}

\subsection{RQ3: Supply Chain Lengths}

\begin{table}
\caption{Distributions of file-level path lengths}
\label{tab:chain}
\begin{tabular}{ccccc}
\toprule
1 & 2 & 3 & 4 & 5 \\
\midrule
150,294 & 12,314 & 1,480 & 30 & 6 \\
\bottomrule
\end{tabular}
\end{table}


Here we consider clone-and-own links as supply chains, and investigate their characteristics.
Figure~\ref{fig:chain} shows the distributions of chain lengths among projects, that is, project-level paths lengths in software supply chain maps. The length of four is the most frequent, and 10 is the longest.
It is revealed that projects are connected directly and indirectly via clone-and-own reusing. Note that since the above mentioned links do not take into account individual files, long chains do not mean the same files are moved to far projects on the chains.

Table~\ref{tab:chain} presents the distributions of file-level path lengths.
We see from this result that how far same files spread in the supply chain maps.
Although most of files are moved only once across projects, some files are found to be transferred on several projects.
This result indicates that the original file authors are sometimes in far ancestor projects.

\begin{tcolorbox}
\textbf{Summary}: In software supply chain maps, projects are connected with clone-and-own supply chains. The project-level chain lengths range from one to ten, and four is most frequent. In file-level, most of files were moved once, and only six files had been moved six times across projects.
\end{tcolorbox}

\section{Discussion}

\subsection{Implications}

Our findings can be summarized into the following implications.
\begin{itemize}
    \item \textit{Many FLOSS projects are connected in a large software supply chain map}. In the studied dataset, we found that many projects form one large map, which indicates that clone-and-own is prevalent and FLOSS projects have complex dependencies among other projects from the point of code sharing.
    \item \textit{Hub projects play an important role in reuse spreading}. As reported in \cite{Gharehyazie:2017:HCC:3104188.3104225}, we also found hub projects that serves as reusing sources for many projects. Such hubs influence consumer projects directly and indirectly, and also have impact on the ecosystem evolution.
    \item \textit{Single repositories are not self-contained}. As we see the transition of file sets across projects, it is clear that tracking original authors within single code repositories is not enough. This fact has big impact on related work and imply the future challenge of building a global version control system for the entire FLOSS ecosystem.
\end{itemize}   

\subsection{Threats to Validity}

Our dataset includes the most popular projects on GitHub.
While the projects include popular applications and libraries according to the number of stars, less popular application projects can also be members of software supply chain maps.
In addition, the number of stars indicates the recent popularity; we might have excluded projects which had been popular before GitHub was launched.
Since excluded projects cannot appear in software supply chain maps, we should have missed existing links in our analysis, and also we might have linked to incorrect projects because of missing appropriate projects. In order to mitigate this threat, we will increase the number of projects to be analyzed.

The result may be affected by programming languages and their ecosystems.
For example, in C language, a common package management system is unavailable.
To use a library, individual developers are required to install the library package provided by an operating system vendor.
To enable a platform-independent build environment, developers have to keep library files in their repository.
On the contrary, JavaScript projects heavily use the npm package management system to reuse functions from existing projects~\cite{Kikas:2017:SEP:3104188.3104203,Decan2018}.
Developers do not need to keep a copy of a library, unless they modified the library files.
Such a practice may change source code reuse activity of developers.

There are limitations in our detection approach.
Since we analyze only tagged snapshots, we could miss the actual reused snapshots in development commits, which are not tagged. In addition, the file similarity threshold can cut file movement histories. Addressing these limitations can be future work.

\section{Related Work}
\subsection{Cross-Project Code Clone Detection}

Code clone detection is the most popular approach to analyze source code reuse activities. 
Since developers may modify a code fragment for their own purpose, various tools have been proposed to detect similar source code fragmetns \cite{Roy2009}.
Kamiya et~al.~\cite{KamiyaTSE2002} proposed CCFinder that analyzes normalized token sequences. 
Jiang et~al.~\cite{Jiang2007} proposed DECKARD that compares a vector representation of an abstract syntax tree of source code. 
Nguyen et~al.~\cite{NguyenFASE2009} proposed  Exas that compares a vector representation  of a program dependence graph.
Sasaki et~al.~\cite{SasakiMSR2010} proposed FCFinder that recognizes file-level clones using hash values of normalized source files.
Cordy et~al.~\cite{Cordy:2011:NCD:2057176.2057234} proposed NiCad that compares a pair of code blocks using a longest common subsequence algorithm.
Sajnani et~al.~\cite{SajnaniICSE2016} proposed SourcererCC that compares a pair of code blocks using Jaccard index of tokens.

Code clone detection tools revealed that software developers often copy source files from other projects.
Hemel et~al.~\cite{Hemel2012} analyzed vendor-specific versions of Linux kernel. 
Their analysis showed that each vendor created a variant of Linux kernel and customized many files in the variant.
Ossher et~al.~\cite{6080795} analyzed cloned files across repositories using a lightweight technique.
They reported that projects cloned files from related projects, libraries, and utilities. 
Koschke et~al.~\cite{KoschkeIWSC2016} also reported that a relatively large number of projects included copies of libraries.
Lopes et~al.~\cite{Lopes:2017:DMC:3152284.3133908} analyzed duplicated files in 4.5 million projects hosted on GitHub and reported that the projects have a large amount of file copies.


While code clone detection tools identify similar files, further analysis is required to identify their origins.
German et~al.~\cite{GermanMSR2009} used CCFinder to detect code siblings reused across FreeBSD, OpenBSD and Linux kernels, 
and then investigated the source code repositories of the projects to identify the original project of a code sibling.
Krinke et~al.~\cite{KrinkeIWSC2010} proposed to distinguish copies from originals by comparing timestamps of code fragments recorded in source code repositories.
Krinke et~al.~\cite{KrinkeMSR2010} used the approach and visualized source code reuse among GNOME Desktop Suite projects.
Our method also uses timestamps recorded in repositories but focuses on a component level code reuse.

Gharehyazie et~al.~\cite{Gharehyazie:2017:HCC:3104188.3104225} analyzed cross-project code clones of 5,753 Java projects on GitHub.  They also analyzed timestamps of the clones and reported that developers often copy an entire library, and some projects serve as hubs (sources) of clones to other projects.
The analyzed source code is a snapshot at a certain point of time.
Our study extends the analysis by including additional programming languages and all the versions of projects.

Ray et~al.~\cite{Ray:2012:CSC:2393596.2393659} analyzed porting of an existing feature or bug fix across forked projects.
They reported that forking allows independent evolution but results in the significant cost of porting activity.
Our maps may help developers to understand the entire history of related projects.

Some researchers consider code reuse across projects as opportunities to extract a reusable component.
Rubin et~al.~\cite{RubinSPLC2013} reported that industrial developers extract reusable components as core assets from existing software products.
Bauer et~al.~\cite{Bauer2013} proposed to extract code clones across products as a candidate of a new library.
Ishihara et~al.~\cite{IshiharaWCRE2013} proposed a function-level clone detection to identify common functions in a number of projects.

Our analysis method assumes that developers simply copy source code without obfuscation tools.
Luo et~al.~\cite{LuoFSE2014} proposed a code plagiarism detection applicable to obfuscated code.
The detection method identifies semantically equivalent basic blocks in two functions.
Chen et~al.~\cite{ChenICSE2014} proposed a technique to detect clones of Android applications using similarity between control-flow graphs of methods.
Those approaches are hard to apply to compare a large set of source files.
Ragkhitwetsagul et~al.~\cite{RagkhitwetsagulSCAM2016} evaluated how the performance of code clone detection and relevant techniques are affected by code obfuscations and optimizations.

Code reuse in binary forms is also out of scope of our method.
S{\ae}bj{\o}rnsen et~al.~\cite{SaebjornsenISSTA2009} proposed a binary clone detection to identify similar code fragments in binaries.
Hemel et~al.~\cite{HemelMSR2011} proposed a binary code clone detection to identify code reuse violating software license of a component.
The method compares the contents of binary files between a target program and each of existing components.
Qiu et~al.~\cite{QiuSANER2015} proposed a code comparison method for a binary form to identify library functions included in an executable file.

For Java software, Davies et~al.~\cite{DaviesMSR2011,DaviesESE2013} proposed a file signature to identify the origin of a jar file using classes and their methods in the file ignoring the details of code. 
German et~al.~\cite{GermanIEEESoftware2012} demonstrated the approach can detect OSS jar files included in proprietary applications.
Mojica et~al.~\cite{MojicaIEEESoftware2014} used the same approach to analyze code reuse among Android applications.
Ishio et~al.~\cite{IshioMSR2016} extended the analysis to automatically identify libraries copied in a product.
Differently from those approaches, our method directly compares source files because small changes in files might be important to understand differences between a cloned component and its original version.


\subsection{Origin Analysis}

Software projects use source code repositories to manage the versions of source code.  
Although a repository tracks modified lines of code between two consecutive versions of a file, the feature is not always sufficient to represent a complicated change.
Godfrey et~al.~\cite{Godfrey2005} proposed origin analysis to identify merged and split functions between two versions of source code.
The method compares identifiers used in functions to identify original functions.
Steidl et~al.~\cite{SteidlMSR2014} proposed to detect source code move, copy, and merge in a source code repository.
The method identifies a similar file in a repository as a candidate of an original version. 

Kawamitsu et~al.~\cite{KawamitsuSCAM2014} proposed an extension of origin analysis across two source code repositories.
Their method identifies an original version of source code in a library's source code repository.
Our method extends the idea to a set of source code repositories.

Spinellis~\cite{Spinellis2016} constructed a Git repository including the entire history of Unix versions.
The unified repository enables developers to investigate the change history of source files across projects.
Since our method visualizes the relationships among projects, construction of a similar unified repository for related projects is a  future direction of our research.

Kanda et~al.~\cite{Kanda2013} proposed a method to recover an evolution history of a product and its variants from their source code archives without a version control.
The approach compares the full contents of source files of products, using a heuristic that developers tend to enhance a derived version and do not often remove code from the derived version.


Antoniol et~al.~\cite{Antoniol2000} proposed a method to recover the traceability links between design documents and source files.  
The method computes similarity of classes by aggregating similarity of their attribute names to identify an original class definition in design documents.
Ishio et~al.~\cite{Ishio:2017:SFS:3104188.3104222} proposed a method to identify the most likely original version of source files by aggregatting file similarity.
Jewmeidang et~al.~\cite{Jewmaidang2018} proposed to visualize how library files are updated in a repository using the method.
Our method also adopted the approach of aggregated similarity to track source code reuse across repositories.

To compare similar software products in detail, several tools have been proposed.
Duszynski~\cite{DuszynskiWCRE2011} proposed a code comparison tool to analyze source code commonalities from a number of similar product variants.
Sakaguchi et~al.~\cite{SakaguchiVISSOFT2015} also proposed a code comparison tool that visualizes a unified directory tree for source files of several products.
Fischer et~al.~\cite{FischerICSME2014} proposed to extract common components from existing product variants and compose a new product.

It should be noted that our analysis method ignores code reuse of a single file or a code fragment.
A typical small-scale reuse is ad-hoc code reuse from the Internet; it has a risk of a license violation~\cite{SojerCACM2011}.
To analyze the problem, Inoue et~al.~\cite{InoueICSE2012} proposed a search engine named Ichi-tracker.  The tool employs existing code search engines and CCFinder to extract similar source code fragments on the Internet.
Yang et~al.~\cite{Yang:2017:SOG:3104188.3104224} employed SourcererCC to identify Python functions on GitHub that are similar to examples on StackOverflow.


\section{Conclusions}

In this study, we defined the concept of software supply chain maps and its construction algorithm.
We built and analyzed the software supply chain maps from 4,592 C/C++/Java projects on GitHub.
The software supply chain maps show that more than a half projects are just consumers, more than a quarter of projects are producers, and a small number of projects are hubs. 
The percentages of the project categories are stable since 2006 to 2016.
The longest software supply chain map includes 10 projects.
The result shows that the clone-and-own source code reuse has been omnipresent in the OSS projects.
To understand the source code of a software project, developers shoudl analyze related projects including the original source files.
The software supply chain maps may help to understand how a project obtained components from existing OSS projects, and how the project is reused by derived projects.


%
\bibliographystyle{ACM-Reference-Format}
\bibliography{ref,references}

\end{document}